\PassOptionsToPackage{unicode}{hyperref}
\PassOptionsToPackage{hyphens}{url}
\PassOptionsToPackage{dvipsnames,svgnames,x11names}{xcolor}
\documentclass[
  11pt,
  twoside]{article}
\usepackage{amsmath,amssymb}
\usepackage{setspace}
\usepackage{iftex}
\ifPDFTeX
  \usepackage[T1]{fontenc}
  \usepackage[utf8]{inputenc}
  \usepackage{textcomp} 
\else 
  \usepackage{unicode-math} 
  \defaultfontfeatures{Scale=MatchLowercase}
  \defaultfontfeatures[\rmfamily]{Ligatures=TeX,Scale=1}
\fi
\usepackage{lmodern}
\ifPDFTeX\else
\fi
\IfFileExists{upquote.sty}{\usepackage{upquote}}{}
\IfFileExists{microtype.sty}{
  \usepackage[]{microtype}
  \UseMicrotypeSet[protrusion]{basicmath} 
}{}
\usepackage{xcolor}
\usepackage[a4paper,top=28mm,bottom=28mm,left=29mm,right=29mm]{geometry}
\usepackage{graphicx}
\makeatletter
\newsavebox\pandoc@box
\newcommand*\pandocbounded[1]{
  \sbox\pandoc@box{#1}%
  \Gscale@div\@tempa{\textheight}{\dimexpr\ht\pandoc@box+\dp\pandoc@box\relax}%
  \Gscale@div\@tempb{\linewidth}{\wd\pandoc@box}%
  \ifdim\@tempb\p@<\@tempa\p@\let\@tempa\@tempb\fi
  \ifdim\@tempa\p@<\p@\scalebox{\@tempa}{\usebox\pandoc@box}%
  \else\usebox{\pandoc@box}%
  \fi%
}
\def\fps@figure{htbp}
\makeatother
\NewDocumentCommand\citeproctext{}{}
\NewDocumentCommand\citeproc{mm}{%
  \begingroup\def\citeproctext{#2}\cite{#1}\endgroup}
\makeatletter
 \let\@cite@ofmt\@firstofone
 \def\@biblabel#1{}
 \def\@cite#1#2{{#1\if@tempswa , #2\fi}}
\makeatother
\newlength{\cslhangindent}
\newlength{\csllabelwidth}
\newenvironment{CSLReferences}[2] 
 {\begin{list}{}{%
  \setlength{\itemindent}{0pt}
  \setlength{\leftmargin}{0pt}
  \setlength{\parsep}{0pt}
  \ifodd #1
   \setlength{\leftmargin}{\cslhangindent}
   \setlength{\itemindent}{-1\cslhangindent}
  \fi
  \setlength{\itemsep}{#2\baselineskip}}}
 {\end{list}}
\usepackage{calc}

\usepackage{hyperref}
\usepackage{BOONDOX-cal}
\usepackage{upgreek}
\usepackage{bm}
\usepackage{fancyhdr}
\usepackage{amsmath}
\usepackage{etoolbox}
\usepackage{nicematrix}
\usepackage[ruled,vlined,linesnumbered]{algorithm2e}
\usepackage[skip=10pt]{caption}
\usepackage{enumitem}

\usepackage{siunitx}
\usepackage[font=footnotesize,labelfont=bf]{caption}
\usepackage[bottom]{footmisc}
\usepackage{amssymb}
\usepackage{tikz}
\usepackage{booktabs}
\usepackage{longtable}
\usepackage{array}
\usepackage{multirow}
\usepackage{wrapfig}
\usepackage{float}
\usepackage{colortbl}
\usepackage{pdflscape}
\usepackage{tabu}
\usepackage{threeparttable}
\usepackage{threeparttablex}
\usepackage[normalem]{ulem}
\usepackage{makecell}
\usepackage{xcolor}
\usepackage{bookmark}
\IfFileExists{xurl.sty}{\usepackage{xurl}}{} 
\hypersetup{
  pdfauthor={Jan H. R. Dressler; Peter Kurz; Winfried J. Steiner},
  colorlinks=true,
  linkcolor={Blue4},
  filecolor={Maroon},
  citecolor={Blue4},
  urlcolor={Blue4},
  pdfcreator={LaTeX via pandoc}}

\title{Computing Nash equilibria for product design\\
based on hierarchical Bayesian mixed logit models}
\author{Jan H. R. Dressler\footnote{\href{https://www.tu-clausthal.de/en/}{Department
  of Marketing, Institute of Management, Economics and Law, Clausthal
  University of Technology, Germany},
  \href{mailto:jhrd13@tu-clausthal.de}{\nolinkurl{jhrd13@tu-clausthal.de}}
  \href{mailto:jan.hendrik.reiner.dressler@tu-clausthal.de}{(corresponding
  author)},
  \href{mailto:winfried.steiner@tu-clausthal.de}{\nolinkurl{winfried.steiner@tu-clausthal.de}}.} \and Peter
Kurz\footnote{\href{https://bms-net.de/en/homepage/}{bms marketing
  research + strategy, Munich, Germany},
  \href{mailto:p.kurz@bms-net.de}{\nolinkurl{p.kurz@bms-net.de}}.} \and Winfried
J. Steiner\footnotemark[1]}
\date{}

\begin{document}
\maketitle
\begin{abstract}
Despite a substantial body of theoretical and empirical research in the
fields of conjoint and discrete choice analysis as well as product line
optimization, relatively few papers focused on the simulation of
subsequent competitive dynamics employing non-cooperative game theory.
Only a fraction of the existing frameworks explored competition on both
product price and design, none of which used fully Bayesian choice
models for simulation. Most crucially, no one has yet assessed the
choice models' ability to uncover the true equilibria, let alone under
different types of choice behavior.

Our analysis of thousands of Nash equilibria, derived in full and
numerically exact on the basis of real prices and costs, provides
evidence that the capability of state-of-the-art mixed logit models to
reveal the true Nash equilibria seems to be primarily contingent upon
the type of choice behavior (probabilistic versus deterministic),
regardless of the number of competing firms, offered products and
features in the market, as well as the degree of preference
heterogeneity and disturbance. Generally, the highest equilibrium
recovery is achieved when applying a deterministic choice rule to
estimated preferences given deterministic choice behavior in reality. It
is especially in the latter setting that incorporating Bayesian
(hyper)parameter uncertainty further enhances the detection rate
compared to posterior means. Additionally, we investigate the influence
of the above factors on other equilibrium characteristics such as
product (line) differentiation.

~


~
\end{abstract}

\setstretch{1}
\renewcommand{\&}{and}
\newcommand{\tikzxmark}{
  \tikz[scale=0.20]{ 
    \draw[line width=0.7,line cap=round](0,0)to[bend left=6](1,1);
    \draw[line width=0.7,line cap=round](0.2,0.95)to[bend right=3](0.8,0.05);
}}
\newcommand{\tikzcmark}{
  \tikz[scale=0.20]{
    \draw[line width=0.7,line cap=round](0.25,0)to[bend left=10](1,1);
    \draw[line width=0.8,line cap=round](0,0.35)to[bend right=1](0.23,0);
}}
\renewcommand{\algorithmautorefname}{Algorithm}

\sisetup{group-separator={,},group-minimum-digits=3} \raggedbottom 
\setlength\evensidemargin{\oddsidemargin} \thispagestyle{empty}
\fancyhf{} \pagestyle{fancy} \setlength{\headheight}{15pt}
\fancyhead[EL]{\thepage} \fancyhead[ER]{\rightmark}
\fancyhead[OR]{\thepage} \fancyhead[OL]{\rightmark}

\subsection{1 Introduction}\label{sec1}

Since the seminal works on conjoint and discrete choice analysis by
\citeproc{ref-1971_Green_Rao}{Green and Rao (1971)} and
\citeproc{ref-1974_McFadden}{McFadden (1974)}, respectively, and their
integration by
\citeproc{ref-1983_Louviere_Woodworth}{Louviere and Woodworth (1983)},
conjoint choice experiments have evolved to a standard technique for
eliciting consumer preferences, often serving as the foundation for
product design in marketing. In such experiments, respondents are shown
multiple sets of alternative product configurations with the task of
selecting their preferred option in each. Assuming that the decisions
are driven by the utilities for the feature levels of the products as
well as unknown influences, suitably formulated random utility models
are then trained on the data to correctly predict the stated (and,
ideally, true) choice behavior through the assignment of choice
probabilities. This opens up the possibility of optimizing a firm's
product (line) given a certain objective and competitive scenario.

Over the past few decades, tremendous progress has been made in the
field of conjoint and discrete choice analysis (extensively shown in,
e.g., \citeproc{ref-2000_Louviere_et_al}{Louviere et al., 2000};
\citeproc{ref-2007_Gustafsson_et_al}{Gustafsson et al., 2007};
\citeproc{ref-2009_Train}{Train, 2009}; \citeproc{ref-2014_Rao}{Rao,
2014}; \citeproc{ref-2015_Hensher_et_al}{Hensher et al., 2015};
\citeproc{ref-2021_Baier_Brusch}{Baier \& Brusch, 2021};
\citeproc{ref-2024_Hess_Daly}{Hess \& Daly, 2024}) as well as product
(line) optimization (see, e.g.,
\citeproc{ref-2008_Belloni_et_al}{Belloni et al., 2008};
\citeproc{ref-2024_Baier_Voekler}{Baier \& Voekler, 2024} for an
overview). Conversely, rather little effort has been devoted to the
game-theoretic simulation of subsequent competitive reactions
(\citeproc{ref-2014_Allenby_et_al}{Allenby et al., 2014};
\citeproc{ref-2021_Steiner_et_al}{Steiner et al., 2021}) employing the
\citeproc{ref-1951_Nash}{Nash (1951)} and
\citeproc{ref-1934_vonStackelberg}{von Stackelberg (1934)} equilibrium
concepts, although competitive scenarios are highly unlikely to remain
constant after the introduction of new or modified products, and firms
are therefore increasingly interested in market equilibria to improve
managerial decisions.

While reflecting on how to attain beneficial insights in this regard,
three main requirements that have to be clarified in advance crossed our
minds. \setlist{nolistsep}

\begin{itemize}[noitemsep]
    \item First and foremost, the estimated choice models must generally be able 
          to uncover the true equilibria, i.e., the equilibria arising under the 
          true consumer preferences. Otherwise, equilibrium simulations would be 
          less valuable. Sufficiently accurate choice predictions do not
          automatically translate into a congruence of the equilibria.
    \item Secondly, considering that choice behavior varies, e.g., depending 
          upon the product category, it is imperative to know whether there
          exist substantial differences between a probabilistic (i.e., logit) 
          and a deterministic (i.e., first) choice rule.
    \item Last but not least, the increased computational cost associated with 
          using the inherent (hyper)parameter uncertainty of a state-of-the-art 
          hierarchical Bayesian mixed logit model for the simulations has to be 
          justified by a notable improvement in the detection of the true 
          equilibria, as opposed to merely an enhancement of the choice 
          predictions.
  \end{itemize}

\noindent

To the best of our knowledge, none of these fundamentally important
aspects seems to have been studied before, neither in the short
(competition on price only) nor in the long run (competition on price
and design, i.e., non-price features). In order to close this research
gap, we conducted a large-scale Monte Carlo study focusing on a firm's
strategic product policy, i.e., the more complex long-run perspective,
and competition between manufacturers directly offering their products
to the consumer. As a basis, we developed a comprehensive system for
conjoint-based game-theoretic simulation (ranging from utility
generation, through choice design and Bayesian model estimation, to Nash
equilibrium computation), which is planned to be published as an R
(\citeproc{ref-2025_RCoreTeam}{R Core Team, 2025}) package in the near
future.

The paper is structured as follows. \hyperref[sec2]{Section 2} briefly
reviews the relevant literature. \hyperref[sec3]{Section 3} covers our
seven-step methodological framework, which allows
\hyperref[sec4]{Section 4} to describe the experimental design in a
concise manner, beginning with the theoretical settings and subsequently
applying them to a realistic use case. \hyperref[sec5]{Section 5}
continues by introducing an overarching scheme of result visuals and
assessing the estimated choice models. Thereupon, statistics
characterizing the equilibria are presented and discussed, divided into
preliminary and pivotal measures. \hyperref[sec6]{Section 6} recaps the
key findings, addresses the study's managerial implications as well as
limitations, and closes with ideas for future research.

\subsection{2 Literature review}\label{sec2}

As can be seen from our compilation of corresponding previous research
in \autoref{tab:literature},
\citeproc{ref-1993_Choi_DeSarbo}{Choi and DeSarbo (1993,}
\citeproc{ref-1994_Choi_DeSarbo}{ 1994)} laid the cornerstone for
conjoint-based simulation of competitive reactions, closely followed by
the works from \citeproc{ref-1995_Gutsche}{Gutsche (1995)},
\citeproc{ref-1997_Green_Krieger}{Green and Krieger (1997)} and
\citeproc{ref-2000_Steiner_Hruschka}{Steiner and Hruschka (2000)}, all
based on the Nash equilibrium concept.
\citeproc{ref-1995_Gutsche}{Gutsche (1995)} was the first to analyze the
effect of the firms' order of movement (see the column titled
\emph{Order}) as well as the effect of their initial product
configurations on the equilibria (column \emph{States}, as we call them
initial states).
\citeproc{ref-2000_Steiner_Hruschka}{Steiner and Hruschka (2000)}
additionally lifted the restriction to the single-product case (column
\emph{Line}). Differentiating between the \emph{Order} and \emph{States}
effect is of importance here because the possible orders of movement of
the firms for each initial state are only covered by starting a Nash
game from every possible initial state if the competitors' objective
functions are strictly symmetric (primarily the prices, cost structure,
consumer preferences and number of products offered). The \emph{Order}
effect is of course absent in methods uncovering the equilibria where
firms do not react sequentially (simultaneous best responses or,
depending on the features' level of measurement, root finding for the
first-order derivatives), but the \emph{States} may always exert an
influence on the equilibrium search if an analytical solution does not
exist.

While some newer papers kept using traditional or thought-up conjoint
data for the Nash (\citeproc{ref-2012_Kuzmanovic_Martic}{Kuzmanovic \&
Martic, 2012}; \citeproc{ref-2017_Liu_et_al}{Liu et al., 2017};
\citeproc{ref-2019_Kuzmanovic_et_al}{Kuzmanovic et al., 2019}) and
Stackelberg games (\citeproc{ref-2010_Steiner}{Steiner, 2010};
\citeproc{ref-2017_Liu_et_al}{Liu et al., 2017}),
\citeproc{ref-2009_Shiau_Michalek}{Shiau and Michalek (2009)},
\citeproc{ref-2011_Wang_et_al}{Wang et al. (2011)} and
\citeproc{ref-2015_Arenoe_et_al}{Arenoe et al. (2015)} started building
on conjoint choice frameworks (column \emph{CBC}) but considered
aggregate or segment demand. With
\citeproc{ref-2012_Chapman_Love}{Chapman and Love (2012)},
\citeproc{ref-2014_Allenby_et_al}{Allenby et al. (2014)},
\citeproc{ref-2019_Hauser_et_al}{Hauser et al. (2019)} and
\citeproc{ref-2021_Bortolomiol_et_al}{Bortolomiol et al. (2021)}, the
literature began to incorporate individual-level heterogeneity (column
\emph{Het.}) through state-of-the-art mixed logit models (column
\emph{MXL}). However, just
\citeproc{ref-2014_Allenby_et_al}{Allenby et al. (2014)} and
\citeproc{ref-2019_Hauser_et_al}{Hauser et al. (2019)} estimated the
latter in a Bayesian fashion and accounted for the (hyper)parameter
uncertainty in their equilibrium computations (column \emph{Draws}),
although not for long-run competition (column \emph{Long}) and product
lines (column \emph{Line}).

\begin{table}[H]
\centering
\caption{\label{tab:literature}Selected works on conjoint-based simulation of competitive reactions}
\centering
\fontsize{7}{9}\selectfont
\begin{threeparttable}
\begin{tabular}[t]{lcccccccccc}
\toprule
Study & CBC & Het. & MXL & Draws & Long & Line & Rules & Order & States & True\\
\midrule
\citeproc{ref-1993_Choi_DeSarbo}{Choi and DeSarbo (1993)} &  & 
  \tikz[scale=0.20]{
    \draw[line width=0.7,line cap=round](0.25,0)to[bend left=10](1,1);
    \draw[line width=0.8,line cap=round](0,0.35)to[bend right=1](0.23,0);
}&  &  & 
  \tikz[scale=0.20]{
    \draw[line width=0.7,line cap=round](0.25,0)to[bend left=10](1,1);
    \draw[line width=0.8,line cap=round](0,0.35)to[bend right=1](0.23,0);
}&  &  &  &  & \\
\citeproc{ref-1994_Choi_DeSarbo}{Choi and DeSarbo (1994)} &  & 
  \tikz[scale=0.20]{
    \draw[line width=0.7,line cap=round](0.25,0)to[bend left=10](1,1);
    \draw[line width=0.8,line cap=round](0,0.35)to[bend right=1](0.23,0);
}&  &  &  &  &  &  &  & \\
\citeproc{ref-1995_Gutsche}{Gutsche (1995) I} &  & 
  \tikz[scale=0.20]{
    \draw[line width=0.7,line cap=round](0.25,0)to[bend left=10](1,1);
    \draw[line width=0.8,line cap=round](0,0.35)to[bend right=1](0.23,0);
}&  &  & 
  \tikz[scale=0.20]{
    \draw[line width=0.7,line cap=round](0.25,0)to[bend left=10](1,1);
    \draw[line width=0.8,line cap=round](0,0.35)to[bend right=1](0.23,0);
}&  &  &  &  & \\
\citeproc{ref-1995_Gutsche}{Gutsche (1995) II} &  & 
  \tikz[scale=0.20]{
    \draw[line width=0.7,line cap=round](0.25,0)to[bend left=10](1,1);
    \draw[line width=0.8,line cap=round](0,0.35)to[bend right=1](0.23,0);
}&  &  &  &  &  & (
  \tikz[scale=0.20]{
    \draw[line width=0.7,line cap=round](0.25,0)to[bend left=10](1,1);
    \draw[line width=0.8,line cap=round](0,0.35)to[bend right=1](0.23,0);
}) & (
  \tikz[scale=0.20]{
    \draw[line width=0.7,line cap=round](0.25,0)to[bend left=10](1,1);
    \draw[line width=0.8,line cap=round](0,0.35)to[bend right=1](0.23,0);
}) & \\
\citeproc{ref-1997_Green_Krieger}{Green and Krieger (1997)} &  & 
  \tikz[scale=0.20]{
    \draw[line width=0.7,line cap=round](0.25,0)to[bend left=10](1,1);
    \draw[line width=0.8,line cap=round](0,0.35)to[bend right=1](0.23,0);
}&  &  & 
  \tikz[scale=0.20]{
    \draw[line width=0.7,line cap=round](0.25,0)to[bend left=10](1,1);
    \draw[line width=0.8,line cap=round](0,0.35)to[bend right=1](0.23,0);
}&  &  & 
  \tikz[scale=0.20]{
    \draw[line width=0.7,line cap=round](0.25,0)to[bend left=10](1,1);
    \draw[line width=0.8,line cap=round](0,0.35)to[bend right=1](0.23,0);
}&  & \\
\citeproc{ref-2000_Steiner_Hruschka}{Steiner and Hruschka (2000)} &  & (
  \tikz[scale=0.20]{
    \draw[line width=0.7,line cap=round](0.25,0)to[bend left=10](1,1);
    \draw[line width=0.8,line cap=round](0,0.35)to[bend right=1](0.23,0);
}) &  &  & 
  \tikz[scale=0.20]{
    \draw[line width=0.7,line cap=round](0.25,0)to[bend left=10](1,1);
    \draw[line width=0.8,line cap=round](0,0.35)to[bend right=1](0.23,0);
}& 
  \tikz[scale=0.20]{
    \draw[line width=0.7,line cap=round](0.25,0)to[bend left=10](1,1);
    \draw[line width=0.8,line cap=round](0,0.35)to[bend right=1](0.23,0);
}&  & 
  \tikz[scale=0.20]{
    \draw[line width=0.7,line cap=round](0.25,0)to[bend left=10](1,1);
    \draw[line width=0.8,line cap=round](0,0.35)to[bend right=1](0.23,0);
}& (
  \tikz[scale=0.20]{
    \draw[line width=0.7,line cap=round](0.25,0)to[bend left=10](1,1);
    \draw[line width=0.8,line cap=round](0,0.35)to[bend right=1](0.23,0);
}) & \\
\citeproc{ref-2009_Shiau_Michalek}{Shiau and Michalek (2009)} & 
  \tikz[scale=0.20]{
    \draw[line width=0.7,line cap=round](0.25,0)to[bend left=10](1,1);
    \draw[line width=0.8,line cap=round](0,0.35)to[bend right=1](0.23,0);
}& (
  \tikz[scale=0.20]{
    \draw[line width=0.7,line cap=round](0.25,0)to[bend left=10](1,1);
    \draw[line width=0.8,line cap=round](0,0.35)to[bend right=1](0.23,0);
}) &  &  &  &  &  &  & (
  \tikz[scale=0.20]{
    \draw[line width=0.7,line cap=round](0.25,0)to[bend left=10](1,1);
    \draw[line width=0.8,line cap=round](0,0.35)to[bend right=1](0.23,0);
}) & \\
\citeproc{ref-2010_Steiner}{Steiner (2010)} &  & (
  \tikz[scale=0.20]{
    \draw[line width=0.7,line cap=round](0.25,0)to[bend left=10](1,1);
    \draw[line width=0.8,line cap=round](0,0.35)to[bend right=1](0.23,0);
}) &  &  & 
  \tikz[scale=0.20]{
    \draw[line width=0.7,line cap=round](0.25,0)to[bend left=10](1,1);
    \draw[line width=0.8,line cap=round](0,0.35)to[bend right=1](0.23,0);
}&  &  & 
  \tikz[scale=0.20]{
    \draw[line width=0.7,line cap=round](0.25,0)to[bend left=10](1,1);
    \draw[line width=0.8,line cap=round](0,0.35)to[bend right=1](0.23,0);
}& 
  \tikz[scale=0.20]{
    \draw[line width=0.7,line cap=round](0.25,0)to[bend left=10](1,1);
    \draw[line width=0.8,line cap=round](0,0.35)to[bend right=1](0.23,0);
}& \\
\citeproc{ref-2011_Wang_et_al}{Wang et al. (2011)} & 
  \tikz[scale=0.20]{
    \draw[line width=0.7,line cap=round](0.25,0)to[bend left=10](1,1);
    \draw[line width=0.8,line cap=round](0,0.35)to[bend right=1](0.23,0);
}& (
  \tikz[scale=0.20]{
    \draw[line width=0.7,line cap=round](0.25,0)to[bend left=10](1,1);
    \draw[line width=0.8,line cap=round](0,0.35)to[bend right=1](0.23,0);
}) &  &  &  &  &  &  &  & \\
\citeproc{ref-2012_Chapman_Love}{Chapman and Love (2012)} & 
  \tikz[scale=0.20]{
    \draw[line width=0.7,line cap=round](0.25,0)to[bend left=10](1,1);
    \draw[line width=0.8,line cap=round](0,0.35)to[bend right=1](0.23,0);
}& 
  \tikz[scale=0.20]{
    \draw[line width=0.7,line cap=round](0.25,0)to[bend left=10](1,1);
    \draw[line width=0.8,line cap=round](0,0.35)to[bend right=1](0.23,0);
}& 
  \tikz[scale=0.20]{
    \draw[line width=0.7,line cap=round](0.25,0)to[bend left=10](1,1);
    \draw[line width=0.8,line cap=round](0,0.35)to[bend right=1](0.23,0);
}&  & (
  \tikz[scale=0.20]{
    \draw[line width=0.7,line cap=round](0.25,0)to[bend left=10](1,1);
    \draw[line width=0.8,line cap=round](0,0.35)to[bend right=1](0.23,0);
}) & 
  \tikz[scale=0.20]{
    \draw[line width=0.7,line cap=round](0.25,0)to[bend left=10](1,1);
    \draw[line width=0.8,line cap=round](0,0.35)to[bend right=1](0.23,0);
}&  &  & 
  \tikz[scale=0.20]{
    \draw[line width=0.7,line cap=round](0.25,0)to[bend left=10](1,1);
    \draw[line width=0.8,line cap=round](0,0.35)to[bend right=1](0.23,0);
}& \\
\citeproc{ref-2012_Kuzmanovic_Martic}{Kuzmanovic and Martic (2012)} &  &  &  &  & 
  \tikz[scale=0.20]{
    \draw[line width=0.7,line cap=round](0.25,0)to[bend left=10](1,1);
    \draw[line width=0.8,line cap=round](0,0.35)to[bend right=1](0.23,0);
}& 
  \tikz[scale=0.20]{
    \draw[line width=0.7,line cap=round](0.25,0)to[bend left=10](1,1);
    \draw[line width=0.8,line cap=round](0,0.35)to[bend right=1](0.23,0);
}&  &  &  & \\
\citeproc{ref-2014_Allenby_et_al}{Allenby et al. (2014)} & 
  \tikz[scale=0.20]{
    \draw[line width=0.7,line cap=round](0.25,0)to[bend left=10](1,1);
    \draw[line width=0.8,line cap=round](0,0.35)to[bend right=1](0.23,0);
}& 
  \tikz[scale=0.20]{
    \draw[line width=0.7,line cap=round](0.25,0)to[bend left=10](1,1);
    \draw[line width=0.8,line cap=round](0,0.35)to[bend right=1](0.23,0);
}& 
  \tikz[scale=0.20]{
    \draw[line width=0.7,line cap=round](0.25,0)to[bend left=10](1,1);
    \draw[line width=0.8,line cap=round](0,0.35)to[bend right=1](0.23,0);
}& 
  \tikz[scale=0.20]{
    \draw[line width=0.7,line cap=round](0.25,0)to[bend left=10](1,1);
    \draw[line width=0.8,line cap=round](0,0.35)to[bend right=1](0.23,0);
}&  &  &  &  &  & \\
\citeproc{ref-2015_Arenoe_et_al}{Arenoe et al. (2015)} & 
  \tikz[scale=0.20]{
    \draw[line width=0.7,line cap=round](0.25,0)to[bend left=10](1,1);
    \draw[line width=0.8,line cap=round](0,0.35)to[bend right=1](0.23,0);
}&  &  &  &  &  &  &  & (
  \tikz[scale=0.20]{
    \draw[line width=0.7,line cap=round](0.25,0)to[bend left=10](1,1);
    \draw[line width=0.8,line cap=round](0,0.35)to[bend right=1](0.23,0);
}) & \\
\citeproc{ref-2017_Liu_et_al}{Liu et al. (2017)} &  & (
  \tikz[scale=0.20]{
    \draw[line width=0.7,line cap=round](0.25,0)to[bend left=10](1,1);
    \draw[line width=0.8,line cap=round](0,0.35)to[bend right=1](0.23,0);
}) &  &  & 
  \tikz[scale=0.20]{
    \draw[line width=0.7,line cap=round](0.25,0)to[bend left=10](1,1);
    \draw[line width=0.8,line cap=round](0,0.35)to[bend right=1](0.23,0);
}& 
  \tikz[scale=0.20]{
    \draw[line width=0.7,line cap=round](0.25,0)to[bend left=10](1,1);
    \draw[line width=0.8,line cap=round](0,0.35)to[bend right=1](0.23,0);
}&  & (
  \tikz[scale=0.20]{
    \draw[line width=0.7,line cap=round](0.25,0)to[bend left=10](1,1);
    \draw[line width=0.8,line cap=round](0,0.35)to[bend right=1](0.23,0);
}) & (
  \tikz[scale=0.20]{
    \draw[line width=0.7,line cap=round](0.25,0)to[bend left=10](1,1);
    \draw[line width=0.8,line cap=round](0,0.35)to[bend right=1](0.23,0);
}) & \\
\citeproc{ref-2019_Hauser_et_al}{Hauser et al. (2019)} & 
  \tikz[scale=0.20]{
    \draw[line width=0.7,line cap=round](0.25,0)to[bend left=10](1,1);
    \draw[line width=0.8,line cap=round](0,0.35)to[bend right=1](0.23,0);
}& 
  \tikz[scale=0.20]{
    \draw[line width=0.7,line cap=round](0.25,0)to[bend left=10](1,1);
    \draw[line width=0.8,line cap=round](0,0.35)to[bend right=1](0.23,0);
}& 
  \tikz[scale=0.20]{
    \draw[line width=0.7,line cap=round](0.25,0)to[bend left=10](1,1);
    \draw[line width=0.8,line cap=round](0,0.35)to[bend right=1](0.23,0);
}& 
  \tikz[scale=0.20]{
    \draw[line width=0.7,line cap=round](0.25,0)to[bend left=10](1,1);
    \draw[line width=0.8,line cap=round](0,0.35)to[bend right=1](0.23,0);
}&  &  &  &  &  & \\
\citeproc{ref-2019_Kuzmanovic_et_al}{Kuzmanovic et al. (2019)} &  & (
  \tikz[scale=0.20]{
    \draw[line width=0.7,line cap=round](0.25,0)to[bend left=10](1,1);
    \draw[line width=0.8,line cap=round](0,0.35)to[bend right=1](0.23,0);
}) &  &  & 
  \tikz[scale=0.20]{
    \draw[line width=0.7,line cap=round](0.25,0)to[bend left=10](1,1);
    \draw[line width=0.8,line cap=round](0,0.35)to[bend right=1](0.23,0);
}& 
  \tikz[scale=0.20]{
    \draw[line width=0.7,line cap=round](0.25,0)to[bend left=10](1,1);
    \draw[line width=0.8,line cap=round](0,0.35)to[bend right=1](0.23,0);
}&  &  & (
  \tikz[scale=0.20]{
    \draw[line width=0.7,line cap=round](0.25,0)to[bend left=10](1,1);
    \draw[line width=0.8,line cap=round](0,0.35)to[bend right=1](0.23,0);
}) & \\
\citeproc{ref-2021_Bortolomiol_et_al}{Bortolomiol et al. (2021) I} & 
  \tikz[scale=0.20]{
    \draw[line width=0.7,line cap=round](0.25,0)to[bend left=10](1,1);
    \draw[line width=0.8,line cap=round](0,0.35)to[bend right=1](0.23,0);
}& 
  \tikz[scale=0.20]{
    \draw[line width=0.7,line cap=round](0.25,0)to[bend left=10](1,1);
    \draw[line width=0.8,line cap=round](0,0.35)to[bend right=1](0.23,0);
}& 
  \tikz[scale=0.20]{
    \draw[line width=0.7,line cap=round](0.25,0)to[bend left=10](1,1);
    \draw[line width=0.8,line cap=round](0,0.35)to[bend right=1](0.23,0);
}&  &  &  &  &  & (
  \tikz[scale=0.20]{
    \draw[line width=0.7,line cap=round](0.25,0)to[bend left=10](1,1);
    \draw[line width=0.8,line cap=round](0,0.35)to[bend right=1](0.23,0);
}) & \\
\citeproc{ref-2021_Bortolomiol_et_al}{Bortolomiol et al. (2021) II} & 
  \tikz[scale=0.20]{
    \draw[line width=0.7,line cap=round](0.25,0)to[bend left=10](1,1);
    \draw[line width=0.8,line cap=round](0,0.35)to[bend right=1](0.23,0);
}& (
  \tikz[scale=0.20]{
    \draw[line width=0.7,line cap=round](0.25,0)to[bend left=10](1,1);
    \draw[line width=0.8,line cap=round](0,0.35)to[bend right=1](0.23,0);
}) &  &  &  & 
  \tikz[scale=0.20]{
    \draw[line width=0.7,line cap=round](0.25,0)to[bend left=10](1,1);
    \draw[line width=0.8,line cap=round](0,0.35)to[bend right=1](0.23,0);
}&  &  & (
  \tikz[scale=0.20]{
    \draw[line width=0.7,line cap=round](0.25,0)to[bend left=10](1,1);
    \draw[line width=0.8,line cap=round](0,0.35)to[bend right=1](0.23,0);
}) & \\
\midrule
This paper & 
  \tikz[scale=0.20]{
    \draw[line width=0.7,line cap=round](0.25,0)to[bend left=10](1,1);
    \draw[line width=0.8,line cap=round](0,0.35)to[bend right=1](0.23,0);
}& 
  \tikz[scale=0.20]{
    \draw[line width=0.7,line cap=round](0.25,0)to[bend left=10](1,1);
    \draw[line width=0.8,line cap=round](0,0.35)to[bend right=1](0.23,0);
}& 
  \tikz[scale=0.20]{
    \draw[line width=0.7,line cap=round](0.25,0)to[bend left=10](1,1);
    \draw[line width=0.8,line cap=round](0,0.35)to[bend right=1](0.23,0);
}& 
  \tikz[scale=0.20]{
    \draw[line width=0.7,line cap=round](0.25,0)to[bend left=10](1,1);
    \draw[line width=0.8,line cap=round](0,0.35)to[bend right=1](0.23,0);
}& 
  \tikz[scale=0.20]{
    \draw[line width=0.7,line cap=round](0.25,0)to[bend left=10](1,1);
    \draw[line width=0.8,line cap=round](0,0.35)to[bend right=1](0.23,0);
}& 
  \tikz[scale=0.20]{
    \draw[line width=0.7,line cap=round](0.25,0)to[bend left=10](1,1);
    \draw[line width=0.8,line cap=round](0,0.35)to[bend right=1](0.23,0);
}& 
  \tikz[scale=0.20]{
    \draw[line width=0.7,line cap=round](0.25,0)to[bend left=10](1,1);
    \draw[line width=0.8,line cap=round](0,0.35)to[bend right=1](0.23,0);
}& 
  \tikz[scale=0.20]{
    \draw[line width=0.7,line cap=round](0.25,0)to[bend left=10](1,1);
    \draw[line width=0.8,line cap=round](0,0.35)to[bend right=1](0.23,0);
}& 
  \tikz[scale=0.20]{
    \draw[line width=0.7,line cap=round](0.25,0)to[bend left=10](1,1);
    \draw[line width=0.8,line cap=round](0,0.35)to[bend right=1](0.23,0);
}& 
  \tikz[scale=0.20]{
    \draw[line width=0.7,line cap=round](0.25,0)to[bend left=10](1,1);
    \draw[line width=0.8,line cap=round](0,0.35)to[bend right=1](0.23,0);
}\\
\bottomrule
\end{tabular}
\begin{tablenotes}[para]
\item \textit{CBC}: building on a choice-based conjoint framework with
            real or simulated respondents, \textit{Het.}: considering preference heterogeneity (in brackets 
            means on segment-level), \textit{MXL}: employing a mixed logit model, \textit{Draws}: accounting for preference uncertainty by using 
            Bayesian (hyper)parameter draws in the simulation of competitive 
            reactions, \textit{Long}: competing on price and design (in brackets means 
            design only), \textit{Line}: optimizing product lines, \textit{Rules}: comparing probabilistic and deterministic choice
            rules, \textit{Order}: incorporating the order of movement effect (in 
            brackets means partly), \textit{States}: starting from every possible initial state, 
            i.e., initial product configuration (in brackets means subspace), \textit{True}: computing the true equilibria as a benchmark in 
            case of a Monte Carlo study.
\end{tablenotes}
\end{threeparttable}
\end{table}

None of the identified papers answered the foundational question if the
models are even capable of revealing the true equilibria (column
\emph{True}), nor did they assess the differences between probabilistic
and deterministic choice rules (column \emph{Rules}, only
\citeproc{ref-1995_Gutsche}{Gutsche, 1995};
\citeproc{ref-2012_Chapman_Love}{Chapman \& Love, 2012} used a
deterministic choice rule), or the benefits of a fully Bayesian model
(over posterior means). As far as we are aware, this also holds true
when expanding from competition among manufacturers to related problems,
e.g., assortment planning (see, e.g.,
\citeproc{ref-2015_Kuxf6k_et_al}{Kök et al., 2015};
\citeproc{ref-2016_Besbes_Saure}{Besbes \& Sauré, 2016}) and combined
manufacturer-retailer settings (see, e.g.,
\citeproc{ref-2007_Luo_et_al}{Luo et al., 2007};
\citeproc{ref-2011_Williams_et_al}{Williams et al., 2011}).

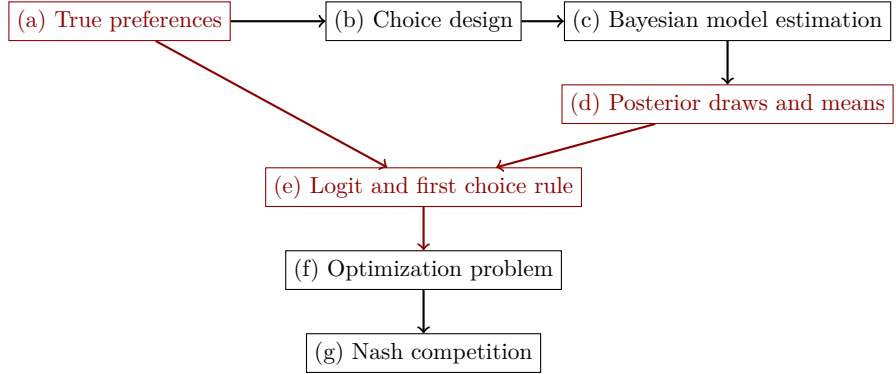
\begin{figure}
  \centering
    \begin{tikzpicture}
      \node (tp) at (0,0)    [draw,Red4,rectangle,scale=0.8] {(a) True preferences};
      \node (cc) at (4,0)    [draw,rectangle,scale=0.8]      {(b) Choice design};
      \node (be) at (8,0)    [draw,rectangle,scale=0.8]      {(c) Bayesian model estimation};
      \node (ep) at (8,-1.1) [draw,Red4,rectangle,scale=0.8] {(d) Posterior draws and means};
      \node (cr) at (4,-2.2) [draw,Red4,rectangle,scale=0.8] {(e) Logit and first choice rule};
      \node (op) at (4,-3.3) [draw,rectangle,scale=0.8]      {(f) Optimization problem};
      \node (nc) at (4,-4.4) [draw,rectangle,scale=0.8]      {(g) Nash competition};
      \draw[->,thick]      (tp)--(cc);
      \draw[->,thick]      (cc)--(be);
      \draw[->,thick]      (be)--(ep);
      \draw[->,thick,Red4] (ep)--(cr);
      \draw[->,thick,Red4] (tp)--(cr);
      \draw[->,thick,Red4] (cr)--(op);
      \draw[->,thick]      (op)--(nc);
    \end{tikzpicture}
  \caption{Visualization of main limitations addressed by our approach}
  \label{lims}
\end{figure}

This paper adds to the existing literature by simulating Nash
competition on price and design between manufacturers directly offering
a product (line) to the consumer, and comparing the resulting equilibria
emerging from hierarchical Bayesian mixed logit estimates (posterior
draws and means) to the ones based on the true consumer preferences,
both for the first and the logit choice rule. The three main limitations
addressed by our approach are summarized by the following diagram
(\autoref{lims}, red elements), which also outlines the methodological
framework we will now dive into. \newline

\subsection{3 Methodological framework}\label{sec3}

The following seven subsections delineate the steps that have been
carried out to ensure computational feasibility and to mitigate any
unintended effects on the outcomes of the simulation study, particularly
those tracing back to insufficient compliance with the currently highest
possible standards and best practices in conjoint choice experiments. We
deliberately refrain from providing specific numerical parameters for
our experimental design to focus solely on the methodological concept
first. The corresponding code is written in R and C++
(\citeproc{ref-2024_Eddelbuettel_et_al}{Eddelbuettel et al., 2024};
\citeproc{ref-2025_RCoreTeam}{R Core Team, 2025}), except for the
generation of choice designs which is done using
SAS\textsuperscript{\textregistered} macros\footnote{\copyright~2024 SAS
  Institute Inc.~SAS and all other SAS Institute Inc.~product or service
  names are registered trademarks or trademarks of SAS Institute Inc.,
  Cary, NC, USA.}. Relevant programming and hardware details are given
within the subsections.

\subsubsection{3.1 Preference generation}\label{sec31}

In a conjoint simulation study it is of utmost importance to generate
synthetic utilities approximating estimates seen in empirical research
and commercial applications. Derived from the latter,
\citeproc{ref-2010a_Wirth}{Wirth (2010a,}
\citeproc{ref-2010b_Wirth}{ 2010b)} proposed an approach for sampling
means and variances of a multivariate normal preference structure that
has recently been verified and employed in thorough simulation studies
conducted by \citeproc{ref-2019_Hein_et_al}{Hein et al. (2019,}
\citeproc{ref-2020_Hein_et_al}{ 2020,}
\citeproc{ref-2022_Hein_et_al}{ 2022)} and
\citeproc{ref-2024_Goeken_et_al}{Goeken et al. (2024)}, building upon
those from \citeproc{ref-1996_Vriens_et_al}{Vriens et al. (1996)},
\citeproc{ref-2002a_Andrews_et_al}{Andrews et al. (2002a,}
\citeproc{ref-2002b_Andrews_et_al}{ 2002b)} and
\citeproc{ref-2003_Andrews_Currim}{Andrews and Currim (2003)}. After
re-examining the resulting densities, again drawing on extensive
practical experience, we decided to closely adhere to this procedure.

Let \(l\in L=\{1,...,\mathcal{l}\}\) be the feature index,
\(m\in M=\{1,...,\mathcal{m}\}\) be the level index,
\(o\in O=\{1,...,\mathcal{o}\}\) be the parameter index such that
\(\mathcal{o}=\mathcal{l}(\mathcal{m}-1)\), and \(\lfloor{\cdot}\rceil\)
be the rounding to the nearest integer function. The definition of
\(\mathcal{o}\) implies modelling of main effects only and having
discrete features (including price) with a constant number of levels,
which are not necessarily prerequisites. However, since firms often only
use certain price points, it seems reasonable to assume discreteness for
the price feature as well. Following
\citeproc{ref-2010a_Wirth}{Wirth (2010a,}
\citeproc{ref-2010b_Wirth}{ 2010b)} we draw
\(\lfloor{0.1\mathcal{o}}\rceil\),
\(\mathcal{o}-2\lfloor{0.1\mathcal{o}}\rceil\),
\(\lfloor{0.1\mathcal{o}}\rceil\) random samples from the continuous
uniform densities \(U(-5,-2)\), \(U(-2,2)\), \(U(2,5)\), respectively,
and concatenate them to a vector \(\bar{\bm{\upbeta}}\) of preference
means. Simultaneously, \(\mathcal{o}\) realizations of a random variable
\(X\) are taken as preference variances, where \begin{gather}
  \begin{gathered}
    \label{prefsimu}
      X=\min(Y+Z_1,Z_2),\\
      Y\sim\Gamma(\kappa,\theta),\\
      Z_1\sim U(\zeta_1,\xi_1),\\
      Z_2\sim U(\zeta_2,\xi_2).
  \end{gathered}
\end{gather} The Gamma density \(\Gamma\), with shape \(\kappa\) and
scale \(\theta\), was fit to empirical data by
\citeproc{ref-2010a_Wirth}{Wirth (2010a,}
\citeproc{ref-2010b_Wirth}{ 2010b)} and truncated using realizations of
the uniformly distributed continuous random variables \(Z_{(\cdot)}\).
Two resulting tuples of distributional parameters allow for a less and a
more heterogeneous structure of preference variances to
emerge.\footnote{\((0.7,1.5,0.08,0.4,9,11)\) and
  \((0.7,4.5,0.2,2,13,18)\) for
  \((\kappa,\theta,\zeta_1,\xi_1,\zeta_2,\xi_2)\).} With the
realizations of \(X\) stacked into a vector \(\bm{\upsigma}^2\), each
pair of components from \(\bar{\bm{\upbeta}}\) and \(\bm{\upsigma}^2\)
is subsequently used to draw a vector \(\bm{\upbeta}_o\) of
\(\mathcal{i}\) random samples from a normal density, where
\(i\in I=\{1,...,\mathcal{i}\}\) denotes the respondent index. This
leads to the preferences being dispersed differently within each feature
level, which is arguably more realistic than setting a constant
variance.

We then choose to assign the vectors \(\bm{\upbeta}_o \forall o\) to the
feature levels by combining them to an \(\mathcal{i}\times\mathcal{o}\)
preference matrix \(\bm{\mathrm{B}}\) according to the order given by
\(\mathcal{o}\) random samples without replacement from a discrete
uniform density \(D(1,\mathcal{o})\). If we wish to define a feature
\(l\) with monotonously changing part-worths corresponding to a specific
sorting of its levels (e.g., price), we actually do the preceding steps
for \(\mathcal{m}\) instead of \(\mathcal{m}-1\) levels of \(l\),
rearrange the \(\mathcal{m}\) entries of \(l\) in \(\bm{\mathrm{B}}\) in
order for each \(i\), shift them so that the first column of \(l\)
becomes a null vector, and lastly remove the latter from
\(\bm{\mathrm{B}}\). This is to avoid any violation of such a
monotonicity condition by the reference category when the coding scheme
of the design matrix is expected to be dummy in the response simulation
and the re-estimation.

\begin{figure}[!ht]

{\centering \includegraphics{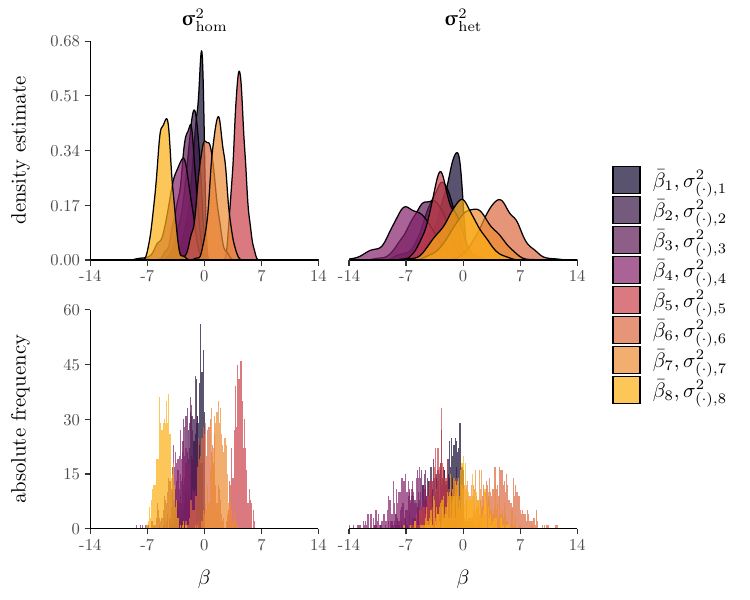} 

}

\caption{Histograms and kernel density estimators of synthetic utilities}\label{fig:synutils}
\end{figure}

\autoref{fig:synutils} gives a typical example of homogeneous and
heterogeneous synthetic utilities generated with the above method. It
displays histograms and kernel density estimators (using Gaussian
kernels and \citeproc{ref-1986_Silverman}{Silverman's, 1986} bandwidth)
for \(\mathcal{l}(\mathcal{m}-1)=2(5-1)=8\) parameters based on
\(\mathcal{i}=500\) individuals, showing plausible differences in
heterogeneity and preference order without violating a monotonicity
constraint exemplarily imposed on \(l=1\) (violet densities). It can
also be seen that the mixtures of normals resulting from this
interference are not a matter of concern.

In regard to the number of respondents to be simulated, we are able to
balance practical experience, theoretical findings as well as
computational and hypothetical financial cost by following the
recommendations of \citeproc{ref-2020_Hein_et_al}{Hein et al. (2020)}
concerning the range. They rigorously analyzed the capabilities of the
hierarchical Bayesian mixed logit model and concluded that under
relatively mild conditions its performance is not significantly affected
by sample size.

\subsubsection{3.2 Choice design}\label{sec32}

With the ongoing objective of preserving generality, we endeavor to
create optimal choice designs for the estimation of main effects from
both the statistical and practical perspective, although artificial
respondents certainly are quite forgiving with respect to some of the
otherwise highly important decisions to be made at this stage.
Therefore, the selection of the numbers of levels, alternatives and
choice sets serves to facilitate meeting respondents' capabilities
(hypothetically) as well as three well-known principles of efficient
choice designs, namely orthogonality, level balance and minimal overlap
(\citeproc{ref-1996_Huber_Zwerina}{Huber \& Zwerina, 1996};
\citeproc{ref-2010_Zwerina_et_al}{Zwerina et al., 2010}). While staying
in the commonly recommended ranges considering cognitive, behavioral and
cost aspects (see, e.g., \citeproc{ref-1996_Johnson_Orme}{Johnson \&
Orme, 1996}; \citeproc{ref-1997_Pinnell_Englert}{Pinnell \& Englert,
1997}; \citeproc{ref-2001_Hensher_et_al}{Hensher et al., 2001};
\citeproc{ref-2005_Caussade_et_al}{Caussade et al., 2005};
\citeproc{ref-2006_Hoogerbrugge_vanderWagt}{Hoogerbrugge \& van der
Wagt, 2006}; \citeproc{ref-2007_Haaijer_Wedel}{Haaijer \& Wedel, 2007};
\citeproc{ref-2012_Kurz_Binner}{Kurz \& Binner, 2012};
\citeproc{ref-2013_Louviere_et_al}{Louviere et al., 2013};
\citeproc{ref-2019_Hein_et_al}{Hein et al., 2019}), we define the number
of levels to be constant across features, the number of alternatives to
be equal to the number of levels and the number of choice sets to be a
multiple of the latter. Symmetric designs also enable us to
(hypothetically) control for a possible number of levels effect
(\citeproc{ref-1981_Currim_et_al}{Currim et al., 1981};
\citeproc{ref-1982_Wittink_et_al}{Wittink et al., 1982},
\citeproc{ref-1990_Wittink_et_al}{1990};
\citeproc{ref-1990_Green_Srinivasan}{Green \& Srinivasan, 1990}).

After these pre-adjustments, we apply the established modified Fedorov
algorithm (\citeproc{ref-1972_Fedorov}{Fedorov, 1972};
\citeproc{ref-1980_Cook_Nachtsheim}{Cook \& Nachtsheim, 1980}) through
the SAS \%ChoicEff macro (\citeproc{ref-2010_Kuhfeld}{Kuhfeld, 2010};
\citeproc{ref-2010_Zwerina_et_al}{Zwerina et al., 2010}) to generate the
designs by maximizing D-efficiency, the most frequently employed
statistical optimality criterion therefor
(\citeproc{ref-2019_Street_Viney}{Street \& Viney, 2019}). As it cannot
be ruled out that the candidate set itself prohibits reaching optimal
D-efficiency if an orthogonal array is used, and because computational
feasibility is given here, we let the candidate set of alternatives from
which the algorithm swaps be the full factorial instead. We always
initiate from multiple (greater than default) random designs
(\citeproc{ref-2010_Kuhfeld}{Kuhfeld, 2010}). By assuming priors of zero
(recall that the Fisher information matrix depends on the parameters in
logit models, \citeproc{ref-2019_Street_Viney}{Street \& Viney, 2019}),
utility balance as the fourth principle is only trivially satisfied
(\citeproc{ref-2010_Zwerina_et_al}{Zwerina et al., 2010}). Motivated
solely by the practical perspective, we additionally enforce the
exclusion of set duplicates.

Other strategies to further improve efficiency, such as specifying
plausible non-zero (Bayesian) priors and blocking larger designs (see,
e.g., \citeproc{ref-2009_Rose_Bliemer}{Rose \& Bliemer, 2009};
\citeproc{ref-2018_Walker_et_al}{Walker et al., 2018}), are deliberately
omitted due to the near-optimal results. Given the artificial nature of
the respondents, there is also no need for randomizing the sets. Designs
for predictive testing (hold-outs) are generated using the same
procedure, with the uniqueness constraint being extended to the union of
training and test sets. Before carrying on, a design assessment
verifying all the required properties is done in R.

\subsubsection{3.3 Error and response simulation}\label{sec33}

Next, following random utility theory, noise must be added to the
individual deterministic portion of utility of the artificial
respondents across alternatives and choice sets prior to the simulation
of choices. For the generation of these random error terms previous
simulation studies fixed a scalar with which to multiply the standard
variance of the chosen error density (see, e.g.,
\citeproc{ref-2002a_Andrews_et_al}{Andrews et al., 2002a};
\citeproc{ref-2003_Andrews_Currim}{Andrews \& Currim, 2003};
\citeproc{ref-2010a_Wirth}{Wirth, 2010a},
\citeproc{ref-2010b_Wirth}{2010b}; \citeproc{ref-2020_Hein_et_al}{Hein
et al., 2020}) or set the variance of the latter to a percentage of the
combined variance of the deterministic and the stochastic portion of
utility (see, e.g., \citeproc{ref-1975_Srinivasan}{Srinivasan, 1975};
\citeproc{ref-1981_Wittink_Cattin}{Wittink \& Cattin, 1981};
\citeproc{ref-1989_Wedel_Steenkamp}{Wedel \& Steenkamp, 1989};
\citeproc{ref-1996_Vriens_et_al}{Vriens et al., 1996};
\citeproc{ref-2002b_Andrews_et_al}{Andrews et al., 2002b}). The second
approach is far more flexible as it automatically adjusts for changes in
the variability of the deterministic utility (e.g., when the number of
features serves as an experimental factor). Nonetheless, explicitly
defining a ratio of measures of dispersion still seems rather intangible
because firstly, the inherent emphasis on larger values has to be taken
into account when using variance or standard deviation. Secondly,
expectation, skewness and kurtosis of both distributions (deterministic
utility and error) are at least computationally neglected, and thirdly,
the stochasticity in the subsequent assignment to the deterministic
utilities is completely unrestricted.

In our view, it is preferable to directly control the proportion of
error to deterministic utility, but to do so, the error has to be
independent from the expectation of the deterministic utilities.
Fortunately, in the case of zero-centering, the deterministic utilities
(as well as the errors) themselves are deviations in their entirety,
which makes the absence of translational invariance irrelevant.
Otherwise, proportionate errors would actually be too large, while not
zero-centering the errors might even make it impossible to sample
proportionate and bidirectional ones. Therefore, zero-centering is a
prerequisite for both.

Based on these thoughts, we propose controlling post hoc for the median
of proportions of error to deterministic utility, which we call the
median relative Gumbel error (MRGE). The distributional choice is
arbitrary and just driven by the nature of the logit model used in this
paper. Let \(\bm{\mathrm{x}}_{jk}\) be a \(1\times \mathcal{o}\) vector
of dummy variables defining the alternative
\(j\in J=\{1,...,\mathcal{j}\}\) in choice set
\(k\in K=\{1,...,\mathcal{k}\}\) such that
\(\bm{\upbeta}_i\bm{\mathrm{x}}_{jk}^T\) is the deterministic utility of
respondent \(i\) for \(j\) in \(k\) (linear-additive part-worth model),
and let \(\varepsilon_{ijk}\) be a realization of the random error
\(E\sim \mathrm{Gumbel}(\lambda,s)\) with location \(\lambda\) and scale
\(s\).\footnote{To draw \(\varepsilon_{ijk}\), we use the R package evd
  (\citeproc{ref-2002_Stephenson}{Stephenson, 2002}).} As usual, an
individual's total utility is assumed to be
\(\bm{\upbeta}_i\bm{\mathrm{x}}_{jk}^T+\varepsilon_{ijk}\). Now, if
\(\bm{\mathrm{v}}_{\mathrm{abs}}\) and
\(\bm{\mathrm{\varepsilon}}_{\mathrm{abs}}\) denote the vectors
containing \(|\bm{\upbeta}_i\bm{\mathrm{x}}_{jk}^T\)\textbar{} and
\(|\varepsilon_{ijk}|\) \(\forall i,j,k\), respectively, and \(\oslash\)
represents the Hadamard division, then the MRGE can be expressed as
\begin{gather}
  \begin{gathered}
    \label{MRGE}
      \mathrm{MRGE}=
        \mathrm{med}(\bm{\mathrm{\varepsilon}}_{\mathrm{abs}}
                     \oslash\bm{\mathrm{v}}_{\mathrm{abs}}).
  \end{gathered}
\end{gather} The MRGE attempts to enhance interpretability by employing
a more accessible fraction (e.g., let the error be 50\% of the
deterministic utility on average) and to increase robustness through its
post hoc inspection. The median is used to prevent the errors from being
scaled down without the suppression of outliers.

Applying the tuning procedure below (\autoref{MRGEtuning}), the standard
deviation \(\sigma\) of the error density is iteratively adjusted until
a required MRGE target value holds. Besides
\(\bm{\mathrm{v}}_{\mathrm{abs}}\), its mean \(\bar{v}_{\mathrm{abs}}\)
and the MRGE target value, it takes a learning rate \(d\), a tolerance
\(t\) and a maximum number of iterations \(r_\mathrm{{max}}\) as inputs,
which do not have to be precisely calibrated to guarantee functionality.

After having generated the errors, the choices are simulated based on
the resulting total utilities and the first choice rule, \begin{gather}
  \begin{gathered}
    \label{fc}
      f_{ijk}=
        \begin{cases}
          1,\,\bm{\upbeta}_i\bm{\mathrm{x}}_{jk}^T+\varepsilon_{ijk}=
              \max_{j'}(\bm{\upbeta}_i\bm{\mathrm{x}}_{j'k}^T+
                \varepsilon_{ij'k}),\\
          0,\,\text{otherwise},
        \end{cases}
      !\exists j'\in J(j'=j).
  \end{gathered}
\end{gather}

\begin{center}
  \begin{minipage}{.58\linewidth}
    \begin{algorithm}[H]
      \footnotesize
        \SetAlCapFnt{\footnotesize}
        \SetAlCapNameFnt{\footnotesize}
        \SetAlgoNlRelativeSize{-2}
      \DontPrintSemicolon
      \SetKwBlock{Repeat}{repeat}{}
        $\bm{\mathrm{input}}$: 
          $\bm{\mathrm{v}}_{\mathrm{abs}}$, $\bar{v}_{\mathrm{abs}}$, 
          $\mathrm{MRGE_{target}}$, $d$, $t$, $r_\mathrm{{max}}$\;
        $h:=\mathrm{MRGE_{target}}$, $r:=1$, $\gamma:=$ Euler's constant\;
        \Repeat{
          $\sigma:=h\cdot\bar{v}_{\mathrm{abs}}$\;
          $s:=\sigma\cdot\frac{\sqrt{6}}{\pi}$\;
          $\lambda:=-s\gamma$\;
          draw $\bm{\mathrm{\varepsilon}}$ from $\mathrm{Gumbel}(\lambda,s)$\;
          $\bm{\mathrm{g}}$ $:=$ 
$\bm{\mathrm{\varepsilon}}_{\mathrm{abs}}\oslash\bm{\mathrm{v}}_{\mathrm{abs}}$\;
          remove undefined quotients from $\bm{\mathrm{g}}$\; 
          $\mathrm{MRGE_{actual}}:=\mathrm{med}(\bm{\mathrm{g}})$\;
          \uIf{$|\mathrm{MRGE_{actual}}-\mathrm{MRGE_{target}}|\le t$}{
            return $\bm{\mathrm{\varepsilon}}$\;
          }\Else{
            $h:=h+(\mathrm{MRGE_{target}}-\mathrm{MRGE_{actual}})\cdot d$\;
            \If{$h<0$}{
              $h:=z$, $Z\sim U(0,1)$\;
            }
          }
          $r:=r+1$\;
          \If{$r>r_{\mathrm{max}}$}{
            return adjust $d$, $t$, $r_{\mathrm{max}}$\;
          }
        }
      \caption{MRGE tuning procedure}\label{MRGEtuning}
    \end{algorithm}
  \end{minipage}
\end{center}

\subsubsection{3.4 Choice model and parameter estimation}\label{sec34}

As stated in \hyperref[sec33]{Section 3.3}, we let the stochastic
portion of utility be i.i.d. Gumbel such that the individual's choice
probability \(\eta_{ijk}\) of choosing alternative \(j\) in choice set
\(k\) given \(\bm{\upbeta}_i\) is defined by the conditional logit model
(\citeproc{ref-1974_McFadden}{McFadden, 1974}) \begin{gather}
  \begin{gathered}
    \label{mnl}
      \eta_{ijk}(\bm{\upbeta}_i)=
        \frac{\mathrm{exp}(\bm{\upbeta}_i\bm{\mathrm{x}}_{jk}^T)}
             {\sum_{j'}\mathrm{exp}(\bm{\upbeta}_i\bm{\mathrm{x}}_{j'k}^T)}.
  \end{gathered}
\end{gather} For the coefficients \(\bm{\upbeta}_i\), a multivariate
normal prior \(\phi\) with \(\mathcal{o}\times1\) mean vector
\(\bar{\bm{\upbeta}}\) and \(\mathcal{o}\times\mathcal{o}\) covariance
matrix \(\bm{\Sigma}\) is specified which describes the population's
unimodal preference heterogeneity and leads to the (unconditional) mixed
logit probability (\citeproc{ref-1999_Brownstone_Train}{Brownstone \&
Train, 1999}) \begin{gather}
  \begin{gathered}
    \label{mlm}
      \mathcal{p}_{ijk}(\bar{\bm{\upbeta}},\bm{\Sigma})=
        \int
          \eta_{ijk}(\bm{\upbeta}_i)
          \phi(\bm{\upbeta}_i|\bar{\bm{\upbeta}},\bm{\Sigma})
        d\bm{\upbeta}_i.
  \end{gathered}
\end{gather} Based on this well-established single component approach
for the distribution of \(\bm{\upbeta}_i\), the likelihood of \(i\)'s
choice sequence \(\bm{\mathrm{y}}_i\) then becomes \begin{gather}
  \begin{gathered}
    \label{likelihood}
      \mathcal{L}(\bm{\mathrm{y}}_i|\bar{\bm{\upbeta}},\bm{\Sigma})=
        \int
          \mathcal{L}(\bm{\mathrm{y}}_i|\bm{\upbeta}_i)
          \phi(\bm{\upbeta}_i|\bar{\bm{\upbeta}},\bm{\Sigma})
        d\bm{\upbeta}_i=
        \int
          \bigg(\prod_{k,j}\eta_{ijk}(\bm{\upbeta}_i)^{f_{ijk}}\bigg)
          \phi(\bm{\upbeta}_i|\bar{\bm{\upbeta}},\bm{\Sigma})
        d\bm{\upbeta}_i.
  \end{gathered}
\end{gather}

For the Bayesian estimation of the hyperparameters
\(\bar{\bm{\upbeta}}\) and \(\bm{\Sigma}\)
(\citeproc{ref-1999_Allenby_Rossi}{Allenby \& Rossi, 1999}), we extend
the hierarchical model by the conjugate hyperprior distributions
\(\phi(\bar{\bm{\upbeta}};\bm{\upmu},\bm{\Omega})\) (normal) and
\(\omega(\bm{\Sigma};\mathcal{v},\bm{\Psi})\) (inverse Wishart) with
\(\mathcal{o}\times1\) mean vector \(\bm{\upmu}\),
\(\mathcal{o}\times\mathcal{o}\) covariance matrix \(\bm{\Omega}\),
degrees of freedom \(\mathcal{v}\) \((\geq\mathcal{o})\) and
\(\mathcal{o}\times\mathcal{o}\) scale matrix \(\bm{\Psi}\). Using
Bayes' theorem and omitting the marginalization over \(\bm{\upbeta}_i\),
the joint posterior of this 3-stage framework can be written as
\begin{gather}
  \begin{gathered}
    \label{posterior}
      P(\bar{\bm{\upbeta}},\bm{\Sigma},\bm{\mathrm{B}}|
        \bm{\mathrm{y}})\propto
          \prod_{i}
            \mathcal{L}(\bm{\mathrm{y}}_i|\bm{\upbeta}_i)
            \phi(\bm{\upbeta}_i|\bar{\bm{\upbeta}},\bm{\Sigma})
            \phi(\bar{\bm{\upbeta}};\bm{\upmu},\bm{\Omega})
            \omega(\bm{\Sigma};\mathcal{v},\bm{\Psi}).
  \end{gathered}
\end{gather} To obtain draws from \(P\), we use the Markov chain Monte
Carlo method (hybrid Gibbs sampler with a random walk
Metropolis-Hastings step) provided by the rhierMnlRwMixture function in
the R package bayesm (\citeproc{ref-2023_Rossi}{Rossi, 2023}), which
allows sampling of \(\bm{\upbeta}_i\forall i\) alongside the
hyperparameters while avoiding simulation of the intractable
\(\mathcal{o}\)-dimensional integral from (\ref{mlm}) and
(\ref{likelihood}) (see \citeproc{ref-2005_Rossi_et_al}{Rossi et al.,
2005}; \citeproc{ref-2009_Train}{Train, 2009} for details). We never
impose constraints on the parameter estimation and always keep the
entire Markov chain for evaluation purposes. However, we control for
monotonicity post hoc if necessary (see \hyperref[sec35]{Section 3.5}).
In the following, we will denote the hyperparameters as \(\Theta\) and
refer to posterior draws of \(\bm{\upbeta}_i\) as just draws.

\subsubsection{3.5 Convergence and model assessment}\label{sec35}

Guided by the recommendations of
\citeproc{ref-2013_Gelman_et_al}{Gelman et al. (2013)} regarding the
diagnosis of convergence, we simulate an independent second chain with
identical settings except for the seed, discard burn-in iterations,
split each chain in half, and assess mixing and stationarity by
computation of the (uni- and multivariate) potential scale reduction
factor (\citeproc{ref-1992_Gelman_Rubin}{Gelman \& Rubin, 1992};
\citeproc{ref-1998_Brooks_Gelman}{Brooks \& Gelman, 1998}) using the
within- and between-sequence variances.\footnote{To compute the
  potential scale reduction factor, we use the R package CODA
  (\citeproc{ref-2006_Plummer_et_al}{Plummer et al., 2006}).} To not
introduce any bias, we reverse-engineer a conservative number of
iterations to hold constant throughout the study for burn-in and
assessment of convergence based on the most demanding experimental
conditions.

When monotonicity constraints are imposed on the synthetic utilities, we
remove violating draws (incorrect order and signs) from the primary
chain, and if the lowest remaining number across individuals is
insufficient, the necessary primary chain length is estimated by
extrapolation. For each individual, the required number of acceptable
draws is then collected starting from the end of the (potentially new)
primary chain. Finally, to reduce serial correlation, we thin the draws
(\citeproc{ref-2009_Train}{Train, 2009};
\citeproc{ref-2013_Gelman_et_al}{Gelman et al., 2013}) independently of
the removal (for the final number of draws to use in simulation, see,
e.g., \citeproc{ref-2000_Orme_Baker}{Orme \& Baker, 2000}).

In line with the Bayesian estimation, the resulting
\(\mathcal{i}\times\mathcal{o}\times\mathcal{n}\) tensor of draws
\(n\in N=\{1,...,\mathcal{n}\}\) is validated by computing credible
intervals for common measures of parameter recovery and predictive
accuracy. If the credibility level \(1-\alpha\) points at decimal
indices for the limits, the weighted mean of the two respective values
of the measure is taken at both limits (i.e., at the lower limit,
\((\frac{\alpha}{2}\mathcal{n})\,\mathrm{mod}\,1\) and
\(1-(\frac{\alpha}{2}\mathcal{n})\,\mathrm{mod}\,1\) as weights for the
lower and upper value, respectively). In the following, we restrict the
representations to the draws, but when deemed necessary, the measures
are also provided for point estimates (posterior means).

To assess the parameter recovery, for each draw we calculate the root
mean squared error (RMSE) between the true and re-estimated individual
part-worths, \(\beta_{io}\) and \(\hat{\beta}_{nio}\), as well as the
average Pearson correlation across the \(\mathcal{i}\) individual
part-worth vector pairs, \(\bm{\upbeta}_{i}\) and
\(\hat{\bm{\upbeta}}_{ni}\) (see, e.g.,
\citeproc{ref-1996_Vriens_et_al}{Vriens et al., 1996};
\citeproc{ref-2002a_Andrews_et_al}{Andrews et al., 2002a},
\citeproc{ref-2002b_Andrews_et_al}{2002b};
\citeproc{ref-2019_Hein_et_al}{Hein et al., 2019},
\citeproc{ref-2020_Hein_et_al}{2020};
\citeproc{ref-2024_Goeken_et_al}{Goeken et al., 2024}). Following
\citeproc{ref-2002a_Andrews_et_al}{Andrews et al. (2002a)}, we assume
that the Gumbel's scale \(s\), known to be inextricably included in
\(\hat{\bm{\upbeta}}_{ni}\) as divisor of the unscaled part-worths, is
estimated correctly as the square root of the factor by which the
standard variance of \(\frac{\pi^2}{6}\) is scaled during the error
generation in the response simulation (see \autoref{MRGEtuning} in
\hyperref[sec33]{Section 3.3}) and can be cancelled out by multiplying
the estimated part-worths with the true \(s\) such that\footnote{To be
  precise, \citeproc{ref-2002a_Andrews_et_al}{Andrews et al. (2002a)}
  introduced the scaling to the simulated part-worths and re-scaled the
  RMSE.} \begin{gather}
  \begin{gathered}
    \label{RMSErec}
      \mathrm{RMSE}_n^{\mathrm{rec}}=
        \sqrt{
          \frac{1}{\mathcal{i}\mathcal{o}}
          \sum_{i,o}\big(s\hat{\beta}_{nio}-\beta_{io}\big)^2
        }.
  \end{gathered}
\end{gather} In regard to the Pearson correlation, which may be the more
appropriate measure of parameter recovery given its scaling invariance,
we think it is quite desirable that the average of coefficients tends to
underestimate the true correlation. Monte Carlo studies also show that
there is a risk of inflating positive bias further by employing the
Fisher transformation for correction (see, e.g.,
\citeproc{ref-2015_Bishara_Hittner}{Bishara \& Hittner, 2015}). In
contrast to previous papers, we therefore calculate the mean
conservatively as \begin{gather}
  \begin{gathered}
    \label{corr}
      \mathrm{corr}_n=
        \frac{1}{\mathcal{i}}
        \sum_{i}\mathrm{corr}\big(\hat{\bm{\upbeta}}_{ni},\bm{\upbeta}_{i}\big).
  \end{gathered}
\end{gather}

The out-of-sample predictive accuracy is evaluated through an individual
as well as an aggregate measure for each draw, namely the hit rate
across hold-out sets averaged over individuals (\autoref{hitrate}) and
the shares of (first) choice RMSE over alternatives and hold-out sets
(\autoref{RMSEsoc}) (see, e.g., \citeproc{ref-1996_Vriens_et_al}{Vriens
et al., 1996}; \citeproc{ref-2002a_Andrews_et_al}{Andrews et al.,
2002a}, \citeproc{ref-2002b_Andrews_et_al}{2002b};
\citeproc{ref-2019_Hein_et_al}{Hein et al., 2019},
\citeproc{ref-2020_Hein_et_al}{2020},
\citeproc{ref-2022_Hein_et_al}{2022};
\citeproc{ref-2024_Goeken_et_al}{Goeken et al., 2024}). \begin{gather}
  \begin{gathered}
    \label{hitrate}
      \mathrm{hitrate}_n=
        \frac{1}{\mathcal{i}}
        \sum_{i}\bigg(\frac{1}{\mathcal{k}}\sum_{k,j}\hat{f}_{nijk}f_{ijk}\bigg)
  \end{gathered}
\end{gather} \begin{gather}
  \begin{gathered}
    \label{RMSEsoc}
      \mathrm{RMSE}_n^{\mathrm{soc}}=
        \sqrt{
          \frac{1}{\mathcal{k}\mathcal{j}}
          \sum_{k,j}
            \bigg(
              \frac{1}{\mathcal{i}}\sum_{i}\hat{f}_{nijk}-
              \frac{1}{\mathcal{i}}\sum_{i}f_{ijk}
            \bigg)^2
        }
  \end{gathered}
\end{gather} The rationale behind the use of the first choice rule here
is rooted in the central part of the methodological framework, i.e., the
next two sections. Given the broader scope of required considerations in
comparison to the logit choice rule, it is just more suitable for the
technical explanations. Hence, analogous to the parameter sets, the
\(\mathrm{RMSE}^{\mathrm{soc}}\) is of course also reported under the
logit rule (\autoref{mnl}) if required.

\subsubsection{3.6 Problem formulation and optimization
method}\label{sec36}

Before undertaking game-theoretic simulations based on the estimates
from \hyperref[sec35]{Section 3.5}, it is necessary to define the
competing firms' objective. Following the seminal conjoint-based
research on non-cooperative competitive reactions from the long-run
perspective (\citeproc{ref-1993_Choi_DeSarbo}{Choi \& DeSarbo, 1993};
\citeproc{ref-1997_Green_Krieger}{Green \& Krieger, 1997}), a firm
\(w\in W=\{1,...,\mathcal{w}\}\) is assumed to search for a product
(line) \(a\in A=\{1,...,\mathcal{a}\}\) that maximizes the total
contribution margin \(\pi_{wak^-}\) given the partial (excluding \(w\))
competitive scenario \(k^-\in K^-=\{1,...,\mathcal{k}^-\}\) by varying
price and design (i.e., non-price feature levels) simultaneously.

To mathematically formulate and extend this optimization problem, also
in view of the few more recent works (recall \autoref{tab:literature}),
let \(J\) now be the set of indices for products in the complete
(including \(w\)) competitive scenario \(k\) (equivalent to a choice set
before), let \(q\in Q=\{1,...,\mathcal{q}\}\) be the product index in
the optimizing firm's line such that \(Q\subset J\), and take
\(|Q|(\geq1)\) as exogeneously fixed. Furthermore, let
\(\bm{\mathrm{p}}\) be a \(1\times\mathcal{m}\) price vector, and let
\(\bm{\mathrm{c}}\) be a \(1\times\mathcal{m}(\mathcal{l}-1)\) vector
that contains the cost of the non-price feature levels. Because the row
vector \(\bm{\mathrm{x}}_{(\cdot)}\) represents a complete product
configuration (see \hyperref[sec33]{Section 3.3}), we are able to
describe price and design separately by splitting
\(\bm{\mathrm{x}}_{(\cdot)}\) into the subvectors
\(\bm{\mathrm{x}}_{(\cdot)p}\) and \(\bm{\mathrm{x}}_{(\cdot)c}\),
respectively. If we expand \(\bm{\mathrm{x}}_{(\cdot)}\) and
\(\hat{\bm{\upbeta}}_{(\cdot)}\) by the reference category of each
feature beforehand, i.e., change the coding scheme of
\(\bm{\mathrm{x}}_{(\cdot)}\) and add zeros to
\(\hat{\bm{\upbeta}}_{(\cdot)}\), the optimization problem can be
written as \begin{align}
  \label{opt1}
    &\!\max_{{\bm{\mathrm{x}}}_{q\in Q}}&& 
      \pi_{wak^-}=\sum_{q}
        \bigg(\frac{1}{\mathcal{n}}\sum_{n,i}\hat{f}_{niq}\bigg)
        \bigg(
          \bm{\mathrm{p}}\bm{\mathrm{x}}_{qp}^T-
          \bm{\mathrm{c}}\bm{\mathrm{x}}_{qc}^T-
          \delta
        \bigg)\\
    \nonumber
      &&&\xrightarrow{a.s.}\sum_{q}
        \bigg(
          \mathcal{i}
            \iint
              f_{q}(\bm{\upbeta})
              \phi(\bm{\upbeta}|\Theta)
              P(\Theta|\bm{\mathrm{y}})
              d\bm{\upbeta}
              d\Theta
        \bigg)
        \bigg(
          \bm{\mathrm{p}}\bm{\mathrm{x}}_{qp}^T-
          \bm{\mathrm{c}}\bm{\mathrm{x}}_{qc}^T-
          \delta
        \bigg)\\
  \label{opt2}
    &\text{s.t.}&&
      \hat{f}_{niq}=
        \begin{cases}
          1,\,\hat{\bm{\upbeta}}_{ni}\bm{\mathrm{x}}_{q}^T=
              \max_{j'}(\hat{\bm{\upbeta}}_{ni}\bm{\mathrm{x}}_{j'}^T)
              \land |S|=1,\\
          \frac{1}{|S|},\,\hat{\bm{\upbeta}}_{ni}\bm{\mathrm{x}}_{q}^T=
                          \max_{j'}(\hat{\bm{\upbeta}}_{ni}
                            \bm{\mathrm{x}}_{j'}^T)
                          \land |S|>1,\\
          0,\,\text{otherwise},
        \end{cases}\\
  \label{opt3}
    &&&
      S=\{j|\hat{\bm{\upbeta}}_{ni}\bm{\mathrm{x}}_{j}^T=
            \textstyle{\max_{j'}}(\hat{\bm{\upbeta}}_{ni}
              \bm{\mathrm{x}}_{j'}^T)\},\\
  \label{opt4}
    &&&
      \bm{\mathrm{x}}_{q^{*}}\bm{\mathrm{x}}^T_q<\mathcal{l}
      \,\mathrm{if}\,|Q|>1,\\
  \label{opt5}
    &&&
      \bm{\mathrm{x}}_q=(\bm{\mathrm{x}}_{qp},\bm{\mathrm{x}}_{qc}),\\
  \label{opt6}
    &&&
      \sum_{m}x_{qlm}=1,\\
  \label{opt7}
    &&&
      x_{qlm}\in \{0,1\}.
\end{align} For each product \(q\), which has to differ from the other
(if existing) products \(q^{*}\in Q(q^*\neq q)\) in at least one feature
(\ref{opt4}), the demand is calculated by taking the sum of first
choices \(\hat{f}_{niq}\) (\ref{opt2}) over individuals \(i\) and
averaging it across draws \(n\). This simulation almost surely converges
to the expected demand, which is the integral over the distribution of
individual preferences \(\bm{\upbeta}\) and the posterior of
hyperparameters \(\Theta\) with regard to the first choices (scaled by
our volume \(\mathcal{i}\) here). Compared with logit probabilities, the
first choices are scaling invariant (i.e., not influenced by the scale
factor \(s^{-1}\) of the part-worths) and immune to the property of
independence of irrelevant alternatives (i.e., not prone to share
inflation for similar products). Their possible downside of unrealistic
determinism (at least for goods with low involvement, see, e.g.,
\citeproc{ref-1979_Shocker_Srinivasan}{Shocker \& Srinivasan, 1979}) is
reduced by implicitly obtaining the individual-level share of draws with
maximum utility for the respective product (note that
\(\frac{1}{\mathcal{n}}\sum_{n,i}\hat{f}_{niq}\) is the same as
\(\sum_{i}(\frac{1}{\mathcal{n}}\sum_{n}\hat{f}_{niq})\)). However,
\(\hat{f}_{niq}\) and \(\hat{\bm{\upbeta}}_{ni}\) will be replaced to
test the effect of various combinations between choice rules (first and
logit choice) and parameter sets (draws vs.~point estimates (posterior
means) vs.~true preferences) on the optimization and equilibria
outcomes. We could as well incorporate the preference uncertainty by
calculating the demand (and optimizing the objective function) for every
draw of the hyperparameters, \begin{gather}
  \begin{gathered}
    \label{optintdraw}
      \sum_{i}\hat{f}_{niq}
      \xrightarrow{a.s.}
      \mathcal{i}\int
        f_{nq}(\bm{\upbeta}_n)
        \phi(\bm{\upbeta}_{n}|\Theta_n)
        d\bm{\upbeta}_{n},
  \end{gathered}
\end{gather} and subsequently simulating the equilibria for every draw
like \citeproc{ref-2014_Allenby_et_al}{Allenby et al. (2014)} to build
up posterior distributions of equilibria. Though, in our case, the
equilibrium quantity is multidimensional (price and non-price features
here vs.~a single metric feature in
\citeproc{ref-2014_Allenby_et_al}{Allenby et al., 2014}), which greatly
restricts the manageability of such distributions with respect to
interpretability and comparability.

If multiple first choices are present within a complete competitive
scenario, indicated by the cardinality of set \(S\), the 100\%
probability is equally divided between them (\ref{opt2}, \ref{opt3}).
Despite being identical with high enough frequency, we intentionally do
not sample because it would produce slightly different total
contribution margins for complete competitive scenarios consisting of
the same set of winning products, which we consider to be unsuitable for
the structural analysis of equilibria (the commonness of recurrences and
ties will become apparent in \hyperref[sec37]{Section 3.7}). It is worth
mentioning that we do not (need to) implement such a tie-breaking
strategy in the response simulation and the model assessment, as a tie
can only arise there in the highly improbable case of two distinct
products showing the identical and at the same time largest total
utility. Furthermore, we do not restrict competition by forcing a lower
bound on the share of choice (cf.
\citeproc{ref-2012_Kuzmanovic_Martic}{Kuzmanovic \& Martic, 2012}).

The expected demand is then multiplied with the corresponding
contribution margin of a single unit, which is computed by subtracting
\(q\)'s cost \(\bm{\mathrm{c}}\bm{\mathrm{x}}_{qc}^T\) as well as a
scalar \(\delta\) from \(q\)'s price
\(\bm{\mathrm{p}}\bm{\mathrm{x}}_{qp}^T\) (and deliberately allowed to
be negative in comparison to, e.g.,
\citeproc{ref-2019_Kuzmanovic_et_al}{Kuzmanovic et al., 2019}). When the
number of features serves as an experimental factor, \(\delta\) does not
only comprise a base cost term but also the cost for features that are
assumed to be unchangeable from the firms' perspective or irrelevant for
the consumers' choice in certain experimental conditions. The marginal
cost of production is assumed to be constant (see also
\citeproc{ref-2014_Allenby_et_al}{Allenby et al., 2014}), as is the cost
of repositioning.

Due to the discrete domain (i.e., set of binary integers) on which the
nonlinear objective function (\ref{opt1}) is defined, the optimization
problem is of combinatorial nature and a solution cannot be derived
analytically. The number of theoretically possible product
configurations \(\tau=\mathcal{m}^{\mathcal{l}}\) and the number of
theoretically possible product line configurations
\(\mathcal{a}=\binom{\tau}{\mathcal{q}}\) grow exponentially with
\(\mathcal{l}\) and \(\mathcal{q}\), respectively. As is the case with
numerous combinatorial optimization problems, there also exists no exact
numeric algorithm capable of solving it in polynomial time, making it
NP-hard (\citeproc{ref-1990_Kohli_Sukumar}{Kohli \& Sukumar, 1990}). In
line with the effort of eliminating uncontrolled systematic influences
on the simulation study, we nevertheless optimize using complete
enumeration such that the structural properties of the equilibria can be
fully captured and are not biased by artefacts from heuristics.

Apart from that, complete enumeration's time complexity does not always
have to be disadvantageous, which is why we also refrain from
implementing other exact methods. Compared with procedures guaranteeing
a solution's global optimality, even the full potential of heuristics
(see \citeproc{ref-2008_Belloni_et_al}{Belloni et al., 2008};
\citeproc{ref-2024_Baier_Voekler}{Baier \& Voekler, 2024} for an
overview) in terms of runtime superiority does not come into play when
the solution space \(A\) of the optimization problem is rather small. In
the simulation of competitive reactions the latter can be
computationally very limited if there is a large number of optimization
problems to be solved consecutively. This will be elaborated upon in the
subsequent section.

\subsubsection{3.7 Pre-computations and Nash competition}\label{sec37}

Given the discrete domain of the objective function, our game-theoretic
solution concept of interest cannot be derived analytically either.
Hence, we simulate dynamic closed-loop games (i.e., multi-stage games
with mutually observable past actions) of myopic best responses in a
sequential manner to obtain the fixed points known as pure strategy Nash
equilibria (\citeproc{ref-1838_Cournot}{Cournot, 1838};
\citeproc{ref-1951_Nash}{Nash, 1951};
\citeproc{ref-1991_Fudenberg_Tirole}{Fudenberg \& Tirole, 1991}).

More precisely, and again closely following
\citeproc{ref-1993_Choi_DeSarbo}{Choi and DeSarbo (1993)} and
\citeproc{ref-1997_Green_Krieger}{Green and Krieger (1997)} as well as
\citeproc{ref-1995_Gutsche}{Gutsche (1995)} and
\citeproc{ref-2000_Steiner_Hruschka}{Steiner and Hruschka (2000)}, firms
take turns in maximizing their total contribution margin \(\pi_{wak^-}\)
depending on the others' product lines until no firm can benefit from
unilaterally changing its product line. In accordance with the
aforementioned articles, the competitors are all assumed to be active
and to be symmetric in regard to prices, cost structure, estimated
consumer preferences and number of products. Moreover, the number of
competitors is expected to remain constant throughout a game (see also
\citeproc{ref-2014_Allenby_et_al}{Allenby et al., 2014}).

If the sequence of each firm optimizing once, \begin{gather}
  \begin{gathered}
    \label{optseq}
      \max_a\pi_{wak^-}\forall w,
  \end{gathered}
\end{gather} is denoted a round \(b \in B=\{1,...,\mathcal{b}\}\) and
\(k^-_0\in K^-\) is the index of the initial state (i.e., the partial
competitive scenario at the beginning of the
\(\mathrm{t\hat{a}tonnement}\)), here, a Nash equilibrium can be
formally expressed as the singleton \begin{gather}
  \begin{gathered}
    \label{nasheq}
      \mathcal{T}^{k^-_0}=
        \{k^{k^-_0}_{b}|k^{k^-_0}_{b}=k^{k^-_0}_{b-1},b\geq 2\}.
  \end{gathered}
\end{gather} \(\mathcal{T}^{k^-_0}\) contains the index
\(k^{k^-_0}_{b}\) referencing the complete competitive scenario \(k\)
which is present at the end of round \(b\) and (first) remained
unchanged for two consecutive rounds after starting from \(k^-_0\). It
is crucial to set an upper limit for the rounds in order to prevent a
game from running infinitely in the absence of an equilibrium,
especially if no comprehensive detection mechanism is implemented that
checks for all different types of cycles. The latter requires the
(partial) comparison between the current \(k_{b}\) and each of the
preceding \(k_{b'}\) with \(b' \in B'=\{1,...,b-2\}\). We decided to
just look for the shortest possible cycle (2-round cycle) and this only
once when the upper limit of the rounds is reached (this is more
efficient if the expected number of cyclic games is rather low because
then this examination is not even triggered once in the majority of
games), \begin{gather}
  \begin{gathered}
    \label{tworoundcycle}
      \mathcal{c}^{k^-_0}_2=
        \begin{cases}
          1,\,k^{k^-_0}_{\mathcal{b}}=k^{k^-_0}_{\mathcal{b}-2},\\
          0,\,\text{otherwise}.
        \end{cases}
  \end{gathered}
\end{gather} On the other hand, at least two rounds have to be played to
see if \(\mathcal{T}^{k^-_0}\neq\emptyset\) because regardless of the
order of movement of the firms, there is no initial state \(k_0\) to
meaningfully compare \(k_1\) with. In other words, a computational
dependence on an initial product line configuration of the firm that
comes first in the reaction sequence does not exist (and we do not
choose one at random), as each game starts with this firm optimizing
over its \(\mathcal{a}\) possibilities.

Thus, there are at most \(\mathcal{k}^-\) initial states
(\(\mathcal{a}^\mathcal{w-1}\) theoretically possible competitive
scenarios of \(\mathcal{w}-1\) firms) to start a game from, which we
exhaustively go through for two reasons. Firstly, the effect of each
unique initial competitive scenario as well as its order of movement
variants (thanks to symmetric competitors) on the equilibria can be
observed, and secondly, every existing equilibrium is guaranteed to be
found since they are inevitably represented in the initial states.

It is noteworthy that with complete enumeration the firm reacting first
already needs to calculate \(\pi_{wak^-}\) once for all \(\mathcal{k}\)
(i.e., \(\mathcal{a}^\mathcal{w}\)) theoretically possible complete
competitive scenarios when optimizing over its \(\mathcal{a}\) product
line configurations at the start of each of the \(\mathcal{k}^-\) games.
If enough memory is available, there are major advantages to
pre-computing a matrix

\AtBeginEnvironment{bNiceMatrix}{\everymath{\displaystyle}}

\begin{gather}
  \begin{gathered}
    \label{m}
      \bm{\mathcal{M}}=
        \begin{bNiceMatrix}
          a_{11}
            &a_{21}
            &\dots
            &a_{\mathcal{w}1}
            &\pi_{11}\\
          \vdots&
            \vdots&
            \ddots&
            \vdots&
            \vdots\\
          a_{1\mathcal{k}}
            &a_{2\mathcal{k}}
            &\dots
            &a_{\mathcal{w}\mathcal{k}}
            &\pi_{1\mathcal{k}}\\
          \CodeAfter
            \OverBrace[yshift=1.5mm,shorten]{1-1}{3-4}
              {\scriptsize{\text{complete competitive scenario}}}
        \end{bNiceMatrix}
      \in\mathbb{R}^{\mathcal{k}\times(\mathcal{w}+1)}
  \end{gathered}
\end{gather} comprising the \(\mathcal{k}\) complete competitive
scenarios as well as the corresponding total contribution margins
\(\pi_{wk}\forall k\) (i.e., \(\pi_{wak^-}\forall a,k^-\)) from an
arbitrary but constant viewpoint \(w\) (e.g., \(w=1\)). After
pre-optimizing \(\mathcal{k}^-\)-times over \(\mathcal{a}\) product line
configurations in \(\bm{\mathcal{M}}\), the resulting matrix

\begin{gather}
  \begin{gathered}
    \label{mopt}
      \bm{\mathcal{M}}^{\mathrm{opt}}=
        \begin{bNiceMatrix}
          \mathop{\mathrm{arg\,max}}_a\pi_{1a1}
            &a_{21}
            &\dots
            &a_{\mathcal{w}1}
            &\max_a\pi_{1a1}\\
          \vdots&
            \vdots&
            \ddots&
            \vdots&
            \vdots\\
          \mathop{\mathrm{arg\,max}}_a\pi_{1a\mathcal{k}^-}
            &a_{2\mathcal{k}^-}
            &\dots
            &a_{\mathcal{w}\mathcal{k}^-}
            &\max_a\pi_{1a\mathcal{k}^-}\\
          \CodeAfter
            \OverBrace[yshift=1.5mm,shorten]{1-2}{3-4}
              {\scriptsize{\text{partial competitive scenario}}}
        \end{bNiceMatrix}
      \in\mathbb{R}^{\mathcal{k}^-\times(\mathcal{w}+1)}
  \end{gathered}
\end{gather} can easily be utilized as a look-up table for the best
response \(\mathop{\mathrm{arg\,max}}_a\pi_{wak^-}\) to a given partial
competitive scenario \(k^-\). Recalling the property of symmetry, it is
evident that \(\bm{\mathcal{M}}^{\mathrm{opt}}\) allows to circumvent
the repeated calculation of identical total contribution margins and
optima across firms and rounds in the games. Consequently, the runtime
of the \(\mathcal{k}^-\) games to be played in an experimental condition
becomes neglectable.

As implicitly stated at the end of \hyperref[sec36]{Section 3.6}, the
\(\mathcal{k}^-\) optimization problems to be solved
(\(\mathcal{k}^-\mathcal{w}\mathcal{b}\) without pre-computation) pose
the main computational bottleneck. Due to the exponential growth with
\(\mathcal{w}\), the solution space \(A\) of a single optimization
problem can be so small that both heuristics and exact methods are
comparably fast but the vast initial state space \(K^-\) precludes
computational feasibility. Thus, if \(A\) calls for heuristics, \(K^-\)
certainly will too. The primary way of restoring feasibility therefore
is to prune \(K^-\). To the best of our knowledge, such an approach has
not yet been developed, but even if it were to exist, it would only be
employed here with the presence of a mathematical proof demonstrating
the equivalence of results.

Apart from the latter, the runtime of the pre-computation of
\(\bm{\mathcal{M}}\) is substantially decreased by implementing the
workhorse functions in C++. Fortunately, the task is embarrassingly
parallel, allowing our three rack servers with 104 physical cores and
2,560 GiB RAM in total to perform at their full computational
capacity.\footnote{ 1x Dell\textsuperscript{\texttrademark}
  PowerEdge\textsuperscript{\texttrademark} R450 with 2x
  Intel\textsuperscript{\textregistered}
  Xeon\textsuperscript{\textregistered} Silver 4316 CPUs and 16x 64 GiB
  DDR4 3200 MT/s RDIMMs, 2x Dell\textsuperscript{\texttrademark}
  PowerEdge\textsuperscript{\texttrademark} R440s with 2x
  Intel\textsuperscript{\textregistered}
  Xeon\textsuperscript{\textregistered} Silver 4216 CPUs and 12x 64 GiB
  DDR4 3200 MT/s RDIMMs each. For reproducible parallel computing, we
  use the R packages doParallel
  (\citeproc{ref-2022a_Microsoft_Weston}{Microsoft Corporation \&
  Weston, 2022a}), foreach
  (\citeproc{ref-2022b_Microsoft_Weston}{Microsoft Corporation \&
  Weston, 2022b}) and doRNG (\citeproc{ref-2025_Gaujoux}{Gaujoux,
  2025}).} Additionally, we pre-compute the possible line and product
configurations of a firm, the \(\tau\times\mathcal{i}\times\mathcal{n}\)
tensor of (exponentiated) product utilities as well as the
\(\tau\times 1\) vector of product contribution margins to serve as
look-ups (\citeproc{ref-2008_Belloni_et_al}{Belloni et al., 2008}). This
avoids unnecessary re-computations also within the main pre-computation
process of \(\bm{\mathcal{M}}\) itself. With regard to memory, the
number of elements in \(\bm{\mathcal{M}}^{({\mathrm{opt}})}\) is
minimized by mapping the pre-computed extended line and product
configurations of each firm to a single integer analogous to (\ref{m}).

Further, we were given the opportunity to test our implementation of the
methodological framework (modified to GPU computing) on a blade server
of an exascale supercomputer currently under development. An
extrapolation of the results indicated that even on entire machines
leading the TOP500 list, simulation of many interesting, yet moderate
scenarios would still be far out of reach, which casts a different light
on the computational limitations in \hyperref[sec4]{Section 4}.

To conclude the methodological framework, we would firstly like to give
reasons for not commenting on a no-choice option in
\hyperref[sec33]{Section 3.3}. Its implementation depends upon the
definition of its \setlist{nolistsep}

\begin{itemize}[noitemsep]
    \item share (assumptions must be made concerning, e.g., the degree of market 
          representation),
    \item attainment (via, e.g., lump-sum after response simulation, calibration 
          of no-choice utility during response simulation) and 
    \item application (only before or also in Nash competition). 
  \end{itemize}

\noindent These variables necessitate its integration as an experimental
factor, which we refrain from doing due to the limited number of
feasible experimental conditions given the computational intensity.
Although it is not a zero-sum game with respect to share anymore when
having a no-choice option to gain from or lose to
(\citeproc{ref-2012_Chapman_Love}{Chapman \& Love, 2012}),
\citeproc{ref-2010_Steiner}{Steiner (2010)} provides evidence that the
no-choice option does not seem to affect the structural properties of
the equilibria. Moreover, none of our measures requires the inclusion of
a no-choice option for interpretation. Secondly,
\hyperref[sec36]{Section 3.6} did not address that the accuracy of the
(relative) cost structure may be another critical factor possibly
affecting the validity of the simulation outcomes when assuming the
objective to be profit maximization (\citeproc{ref-1990_Choi_et_al}{Choi
et al., 1990}; \citeproc{ref-1994_Choi_DeSarbo}{Choi \& DeSarbo, 1994}).
Fortunately, we are in the comfortable position of having had access to
real costs. Details will be given in the upcoming section.

\subsection{4 Design of the simulation study}\label{sec4}

\subsubsection{4.1 Theoretical settings}\label{sec41}

As the feasible number of experimental conditions is heavily bounded by
the computational complexity, we decided to hold the number of
respondents (\(\mathcal{i}=500\)), levels (\(\mathcal{m}=5\)),
alternatives per choice set (\(\mathcal{j}=5\)), choice sets
(\(\mathcal{k}_{\mathrm{train}}=15\))\footnote{The D-efficiency of the
  generated choice designs was between 96.8\% and 99.6\% with a prior of
  zero and 50 random start designs.} and hold-out sets
(\(\mathcal{k}_{\mathrm{test}}=5\)) constant throughout the simulation
study. We always imposed a monotonicity constraint on the first feature
when generating the preferences, assuming it to represent price without
being an indicator for, e.g., quality. Following the findings of
\citeproc{ref-2019_Hein_et_al}{Hein et al (2019)}, the prior
specifications for model estimation were left at default. We also fixed
the number of iterations for burn-in and assessment of convergence
(10,000 and 30,000, respectively). Note that the removal of violating
draws forced the chains to be much longer (median \(>\) 120,000).
Lastly, we chose to anchor the thinning factor (10), the number of draws
to be used in subsequent computations (\(\mathcal{n}=500\)) and the
upper limit of the game rounds (\(\mathcal{b}=20\)) as well.

We then calculated the number of complete competitive scenarios
(\(\mathcal{k}\)) for systematically varied numbers of features
(\(\mathcal{l}\)), products (\(\mathcal{q}\)) and firms
(\(\mathcal{w}\)), measured the required runtime for small instances of
\(\bm{\mathcal{M}}\) (i.e., \(\mathcal{k}\)) and extrapolated it to the
other cases (proven to be reliable estimates). After ordering the
resulting table by the extrapolated runtime, it was cut off at the point
where the latter increased from less than three days to roughly a month
for the next fastest condition (\(\mathcal{k}\approx 3\) billion). Keep
in mind that disregarding hyperparameter uncertainty reduces runtime by
a factor close to \(\mathcal{n}\). In light of the fact that the
truncation interferes with the perfect systematic variation of the above
mentioned variables controlling \(\mathcal{k}\), they will be referred
to as imperfect experimental factors (\autoref{tab:expfacs}).

\begin{table}[H]
\centering
\caption{\label{tab:expfacs}Experimental factors}
\centering
\fontsize{9}{11}\selectfont
\begin{tabular}[t]{ll}
\toprule
Factor & Levels\\
\midrule
\addlinespace[0.3em]
\multicolumn{2}{l}{\textit{Imperfect}}\\
\hspace{1em}\#Features & 2, 3, 4, 5, 6\\
\hspace{1em}\#Products & 1, 2, 3, 4\\
\hspace{1em}\#Firms & 2, 3, 4, 5\\
\addlinespace[0.3em]
\multicolumn{2}{l}{\textit{Perfect}}\\
\hspace{1em}Preference structure & Hom., het.\\
\hspace{1em}Error magnitude & 12.5\%, 50\%\\
\hspace{1em}Choice rule & First, logit\\
\hspace{1em}Parameter set & Draws, means, true\\
\bottomrule
\multicolumn{2}{l}{\rule{0pt}{1em}Replications: 3 (unique sets of seeds)}\\
\end{tabular}
\end{table}

For the remaining 16 base conditions (\autoref{tab:basecon}) defined by
the three imperfect experimental factors, we specified four perfect
experimental factors (\autoref{tab:expfacs}), namely the preference
structure (homogeneous vs.~heterogeneous variances), the error magnitude
(small (12.5\%) vs.~large (50\%) MRGE\footnote{The parameter settings
  for the MRGE tuning procedure were: \(d=0.5\), \(t=10^{-5}\),
  \(r_{\mathrm{max}}=10^{4}\).}), the choice rule (first vs.~logit
choice) and the parameter set (draws vs.~point estimates (posterior
means) vs.~true preferences), leading to \(16\times 2^2=64\)
experimental conditions with \(2 \times 3=6\) choice rule \(\times\)
parameter set combinations each. To eliminate the influence of the
stochastic processes involved, each of the 64 experimental conditions
was run with a unique set of seeds and replicated three times with
another three unique sets of seeds, resulting in 256 experimental
conditions differing in seed.

\begin{table}[H]
\centering
\caption{\label{tab:basecon}Base conditions}
\centering
\fontsize{9}{11}\selectfont
\begin{tabular}[t]{rcccrrr}
\toprule
\multicolumn{1}{c}{ } & \multicolumn{3}{c}{Imperfect experimental factors} \\
\cmidrule(l{3pt}r{3pt}){2-4}
 & \makecell[c]{\#Features \\ ($\mathcal{l}$)} & \makecell[c]{\#Products \\ ($\mathcal{q}$)} & \makecell[c]{\#Firms \\ ($\mathcal{w}$)} & \makecell[c]{\#Product \\ config. ($\tau$)} & \makecell[c]{\#Line \\ config. ($\mathcal{a}$)} & \makecell[c]{\#Competitive \\ scenarios ($\mathcal{k}$)}\\
\midrule
1 & 2 & 1 & 2 & \num{25} & \num{25} & \num{625}\\
2 & 3 & 1 & 2 & \num{125} & \num{125} & \num{15625}\\
3 & 2 & 1 & 3 & \num{25} & \num{25} & \num{15625}\\
4 & 2 & 2 & 2 & \num{25} & \num{300} & \num{90000}\\
5 & 4 & 1 & 2 & \num{625} & \num{625} & \num{390625}\\
6 & 2 & 1 & 4 & \num{25} & \num{25} & \num{390625}\\
7 & 3 & 1 & 3 & \num{125} & \num{125} & \num{1953125}\\
8 & 2 & 3 & 2 & \num{25} & \num{2300} & \num{5290000}\\
9 & 5 & 1 & 2 & \num{3125} & \num{3125} & \num{9765625}\\
10 & 2 & 1 & 5 & \num{25} & \num{25} & \num{9765625}\\
11 & 2 & 2 & 3 & \num{25} & \num{300} & \num{27000000}\\
12 & 3 & 2 & 2 & \num{125} & \num{7750} & \num{60062500}\\
13 & 2 & 4 & 2 & \num{25} & \num{12650} & \num{160022500}\\
14 & 6 & 1 & 2 & \num{15625} & \num{15625} & \num{244140625}\\
15 & 4 & 1 & 3 & \num{625} & \num{625} & \num{244140625}\\
16 & 3 & 1 & 4 & \num{125} & \num{125} & \num{244140625}\\
\bottomrule
\end{tabular}
\end{table}

Whenever less than the maximum \#features was present, the excluded
features were kept out from the beginning (i.e., the conjoint choice
experiment), and we randomly chose their levels to add the corresponding
costs to the base cost \(\delta\) (see \autoref{tab:featmat} in the
following section).

\subsubsection{4.2 Use case settings}\label{sec42}

In line with our objective of obtaining valid outcomes by conducting a
simulation study resembling real-world conditions whenever possible, we
applied the generic settings from \hyperref[sec41]{Section 4.1} to the
tangible example of notebooks and, most importantly, were able to gain
insight into the (relative) cost structure of a well-known firm in this
sector, solving the difficulty mentioned at the end of
\hyperref[sec37]{Section 3.7}. We therefore model competition between
multinational computer manufacturers who offer their products directly
to the consumers.

To do so, we determined five modifiable, discriminating features driving
the consumers' choice in addition to price, namely display size, central
processing unit (CPU), solid state drive (SSD) capacity, battery life
and random access memory (RAM). Their mutually exclusive levels were
chosen to cover the range of existing possibilities. The features,
levels and corresponding costs (including base costs) are displayed in
\autoref{tab:featmat}. If features were removed or added for a base
condition, this was done according to the order given by
\autoref{tab:featmat}.

\begin{table}[H]
\centering
\caption{\label{tab:featmat}Notebook features, levels and costs}
\centering
\fontsize{9}{11}\selectfont
\begin{tabular}[t]{>{\raggedright\arraybackslash}p{2cm}>{\raggedleft\arraybackslash}p{2cm}>{\raggedleft\arraybackslash}p{2cm}>{\raggedleft\arraybackslash}p{2cm}>{\raggedleft\arraybackslash}p{2cm}>{\raggedleft\arraybackslash}p{2cm}}
\toprule
  & Level 1 & Level 2 & Level 3 & Level 4 & Level 5\\
\midrule
\midrule
Price & \num{299} € & \num{599} € & \num{899} € & \num{1199} € & \num{1499} €\\
\midrule
Display size & 13 " & 14 " & 15 " & 16 " & 17 "\\
 & 25 € & 30 € & 33 € & 44 € & 54 €\\
\midrule
CPU & *i3 & **Ryzen\textsuperscript{\texttrademark} 5 & *i5 & **Ryzen\textsuperscript{\texttrademark} 7 & *i7\\
 & 10 € & 11 € & 12 € & 65 € & 79 €\\
\midrule
SSD capacity & \num{125} GB & \num{250} GB & \num{500} GB & \num{1000} GB & \num{2000} GB\\
 & 11 € & 11 € & 11 € & 23 € & 31 €\\
\midrule
Battery life & 5 h & 7 h & 9 h & 11 h & 13 h\\
 & 8 € & 8 € & 10 € & 10 € & 12 €\\
\midrule
RAM & 4 GB & 8 GB & 16 GB & 32 GB & 64 GB\\
 & 6 € & 6 € & 9 € & 19 € & 38 €\\
\bottomrule
\multicolumn{6}{l}{\rule{0pt}{1em}Base cost (e.g., housing, mainboard, keyboard, touchpad): 94 €}\\
\multicolumn{6}{l}{\rule{0pt}{1em}*Intel\textsuperscript{\textregistered} Core\textsuperscript{\texttrademark}, **AMD}\\
\end{tabular}
\end{table}

\subsection{5 Results and discussion}\label{sec5}

Since the estimated choice models are fundamental to our game-theoretic
simulations, \hyperref[sec5]{Section 5} begins with briefly presenting
the outcomes of their assessment described in \hyperref[sec35]{Section
3.5}. Prior to that, however, it seems appropriate to devote a few
general sentences to the corresponding visuals because they will
accompany the reader throughout the remainder of the section as well,
i.e., the preliminary and pivotal equilibrium measures.\footnote{All the
  figures were build with the R package ggplot2
  (\citeproc{ref-2016_Wickham}{Wickham, 2016}).}

\subsubsection{5.1 Visuals}\label{sec51}

\begin{figure}[!ht]

{\centering \includegraphics{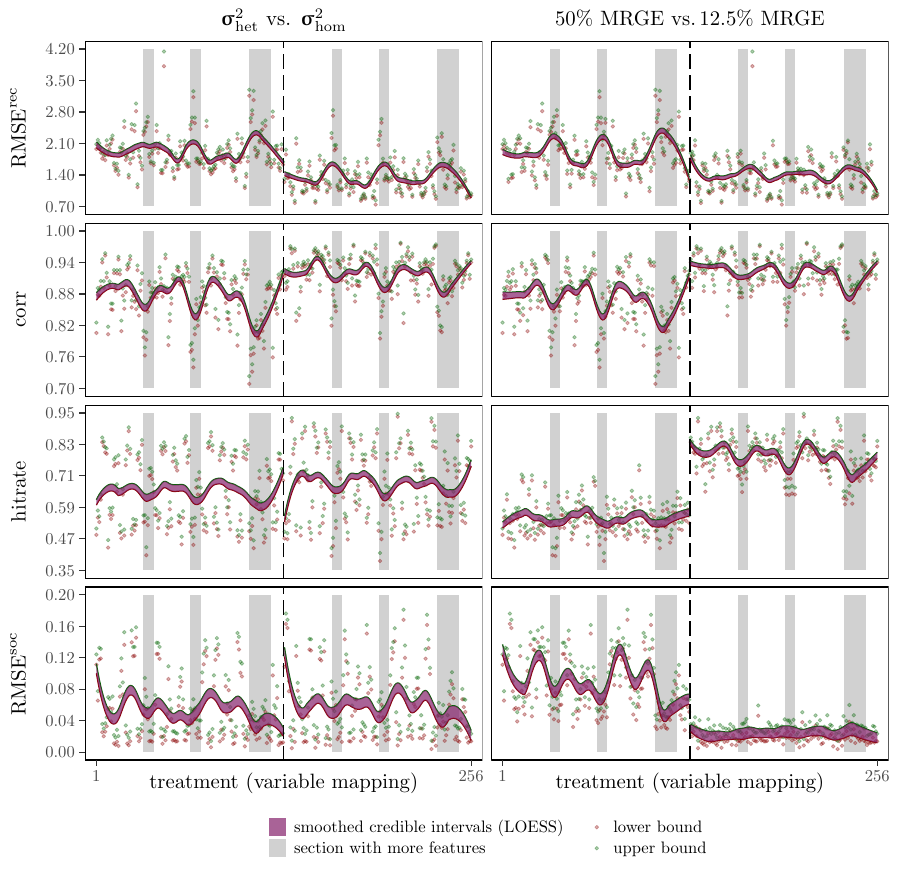} 

}

\caption{$95\%$ credible intervals for measures of parameter  recovery and predictive accuracy}\label{fig:accuracy}
\end{figure}

In order to enable the display of fine details without compromising
interpretability, particularly in view of the equilibrium measures,
three main aspects have been thought of, which can initially be seen in
\autoref{fig:accuracy}. First, we stick to a single scheme. Once
explained, subsequent findings are much easier to grasp. Second, we plot
the data as disaggregated as possible and use smoothing techniques to
highlight trends. Third, we include the imperfect experimental factors
in a subtle manner, also doing justice to their missing systematic
variation.

Each row belongs to a measure and each column to a level of the perfect
experimental factors preference structure and error magnitude. Each
column therefore contains 128 of the 256 experimental conditions
(treatments) differing in seed. In each column, the x-axis is sorted in
such a way that for each of the 16 base conditions (in the order given
by \autoref{tab:basecon}) there are 4 runs differing in seed for each of
the 2 levels of the other perfect experimental factor (preference
structure or error magnitude, in the order given by the column headers).
The grey shaded areas mark the sections with higher levels of the
imperfect experimental factors. Higher levels means \#features \(\geq4\)
(cf.~\autoref{tab:basecon}, rows 5, 9, 14-15), \#products \(\geq2\)
(cf.~rows 4, 8, 11-13) and \#firms \(\geq4\) (cf.~rows 6, 10, 16),
resulting in three grey bars each and different widths of the bars. As
the imperfect experimental factors \#products and \#firms are irrelevant
for the measures in \autoref{fig:accuracy} (together with the perfect
experimental factors choice rule and parameter set for now), just three
bars in a single shade of grey are visible, referring to the higher
\#features (i.e., model complexity). Either local regression,
specifically LOESS, or a simple moving average (SMA) is used for
smoothing the scatter plots.

\subsubsection{5.2 Model assessment}\label{sec52}

For each experimental condition, \autoref{fig:accuracy} shows the lower
(red dots) and the upper (green dots) \(95\%\) credible bound of the
measures of parameter recovery (\(\mathrm{RMSE}^{\mathrm{rec}}\),
\(\mathrm{corr}\)) and predictive accuracy (\(\mathrm{hitrate}\),
\(\mathrm{RMSE}^{\mathrm{soc}}\)) introduced in \hyperref[sec35]{Section
3.5}. The purple band indicates the span of the bounds smoothed by local
regression.

As anticipated, all measures clearly benefit from less disturbance
(cf.~right-hand columns). The same holds true for a lesser \#features
(white sections) across both perfect experimental factors (preference
structure and error magnitude) and their levels, with the exception of
the \(\mathrm{RMSE}^{\mathrm{soc}}\) (last row). Interestingly, the
latter tends to improve with a higher \#features. Also not unexpectedly,
greater homogeneity enhances parameter recovery (upper left quadrant of
\autoref{fig:accuracy}). The predictive accuracy, however, is only
marginally influenced by the preference structure (bottom left quadrant
of \autoref{fig:accuracy}). To us, the two slightly less intuitive
findings are not of any concern, as they have been observed to some
extent in previous simulation studies (see, e.g.,
\citeproc{ref-2002b_Andrews_et_al}{Andrews et al., 2002b};
\citeproc{ref-2020_Hein_et_al}{Hein et al., 2020}).

\subsubsection{5.3 Equilibria}\label{sec53}

In the visuals for the following central part of the results, as first
illustrated by \autoref{fig:prelimeqmeasures}, each of the 128
experimental conditions per column principally comprises six data points
in six different colors for the six choice rule \(\times\) parameter set
combinations. The six lines represent the six corresponding simple
moving averages. Two color schemes are employed to be able to easily
differentiate between the two choice rules (e.g., green-yellow scheme
for the three logit choice rule combinations).

With regard to the equilibria, identical ones are only counted once per
run. In each of the \(256 \times 6 =\) 1,536 total runs, between 25 and
1,953,125 games had to be simulated
(\(\mathcal{k}^-=\mathcal{a}^{\mathcal{w}-1}\) initial states, see
\autoref{tab:basecon}), which of course often led to identical
equilibria multiple times in a run. After eliminating these duplicates,
totals of 4,638 and 2,921 equilibria with and without flips remain for
analysis, respectively. Flips of an equilibrium consist of the same
products or lines but swapped between firms (recall the competitors'
symmetry). Furthermore, and most crucially, let us assume that the true
equilibria are determinable by means of the simulated preferences. This
allows us to assess the models' ability to reveal the truth, which to
the best of our knowledge has not yet been done. Both the preliminary
and the pivotal measures reported in \hyperref[sec531]{Section 5.3.1}
and \hyperref[sec532]{Section 5.3.2}, respectively, serve the purpose of
exploring the influence of the experimental factors on the equilibria,
but the pivotal measures are more indicative of the validity and
stability of the equilibria based on the estimated parameter sets.

\subsubsection{5.3.1 Preliminary measures}\label{sec531}

\begin{figure}[!ht]

{\centering \includegraphics{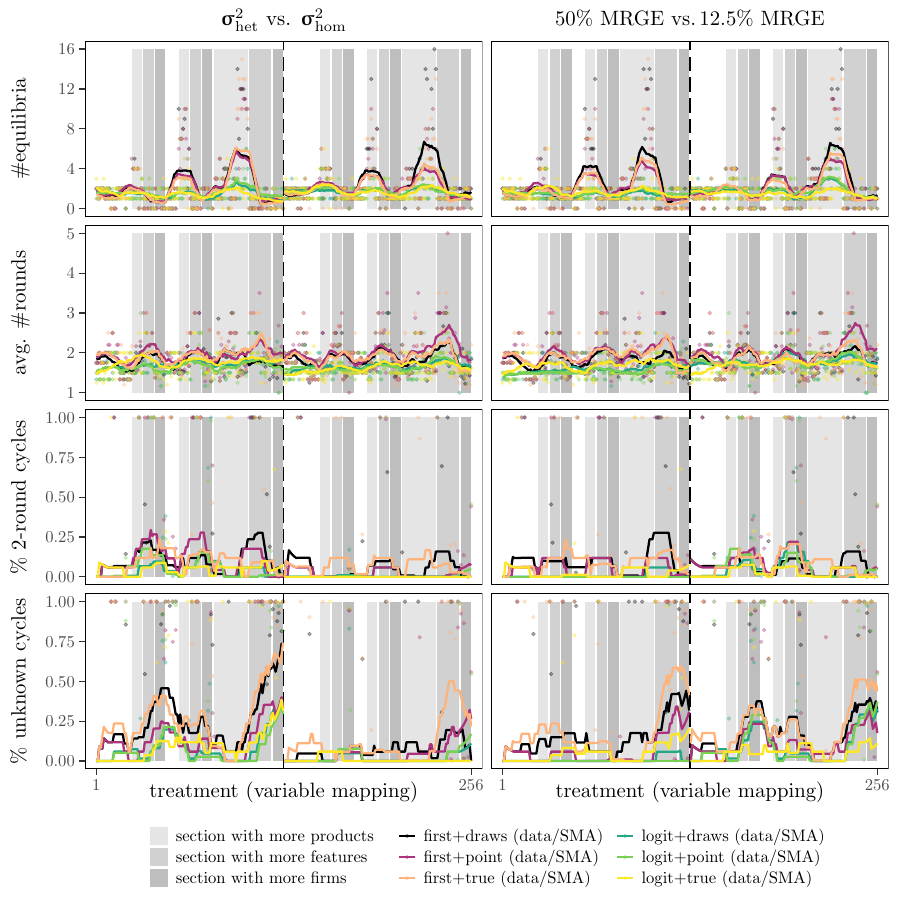} 

}

\caption{Preliminary equilibrium measures}\label{fig:prelimeqmeasures}
\end{figure}

The first of the four preliminary measures displayed in
\autoref{fig:prelimeqmeasures} is the number of equilibria. Flips are
excluded to ensure that it is not biased by the \#firms and their
differentiation because a differentiated equilibrium can be flipped more
often with a growing number of competitors, whereas an undifferentiated
one (where all firms offer the identical product or lines) cannot be
flipped at all. While the impact of the preference structure, error
magnitude and parameter set (draws, point estimates, true preferences)
seems negligible if there exist equilibria (\(y\neq 0\), 87.6\% of all
runs), the number of first choice equilibria rises noticably with the
\#products (from \(\leq 4\) to \(\leq 16\)). An explanation could be
that compared to the logit choice rule an infinitesimal tendency in
preferences is theoretically sufficient to substantially shift the
distribution of shares across alternatives, which might increase the
variance of mutual best responses. Simultaneously, the first choice rule
is usually responsible for not reaching an equilibrium from any of the
initial states, predominantly in the single product cases (see local
minima of SMA (simple moving averages) and \(y=0\)). Within each choice
rule, both the draws and the posterior means are able to simulate the
true numbers quite accurately (cf.~SMA). Last but not least, the
likelihood of finding an equilibrium is definitely boosted by
homogeneity (see \(y=0\) in column \(\bm{\upsigma}^2_\mathrm{hom}\)).

Next, the average number of rounds necessary to reach these equilibria
is shown. Flips are included to enlarge the data base because they do
not skew this measure. Except for the spikes occurring in sections with
a higher \#features instead of \#products (also for the true parameters,
so it cannot be related to changes in model performance due to
complexity), the pattern is quite similar to the one above. Though,
given the small range of rounds (mainly \(1\leq b\leq 3\)), there is not
much to conclude here besides that the first choice rule takes a
fraction of a round longer on average.

The last two preliminary measures are closely linked. If a game in a run
does not end in an equilibrium, either more rounds have to be played or
a cycle is present. The measure \% 2-round cycles is the number of
2-round cycles (see \hyperref[sec37]{Section 3.7}) divided by the number
of games (i.e., initial states) simulated in the respective run. The
measure \% unknown cycles technically catches all other cases, which, in
our view, means all other cycles. As written in \hyperref[sec37]{Section
3.7}, we are only certain about the 2-round cycles, but the upper limit
of rounds (\(\mathcal{b}=20\)) being too low seems very unlikely given
that the average number of rounds never exceeds five. Whenever both
percentages sum to one, none of the games in the run led to an
equilibrium (\(y=0\) in \#equilibria, 12.4\% of all runs). The runs with
0\% 2-round or unknown cycles (91.6\% and 88.5\%, respectively) are
visually hidden to not cover the SMA close to the x-axis. Apart from the
dominance of the first choice rule and the effect of the preference
structure already mentioned in connection with the \#equilibria, no
obvious regularities can be inferred from the experimental factors under
consideration with respect to these two measures. This might be due to
fewer observations (only 8.4\% and 11.5\% of runs show 2-round and
unknown cycles, respectively), which is supported by the fact that there
is not only a lesser degree of congruence between the shares based on
the true and the estimated parameter sets but also between the shares
based on the latter (draws and point estimates). At best, one could note
the recurring peaks in sections of a higher \#features or \#firms.

\subsubsection{5.3.2 Pivotal measures}\label{sec532}

\begin{figure}[!ht]

{\centering \includegraphics{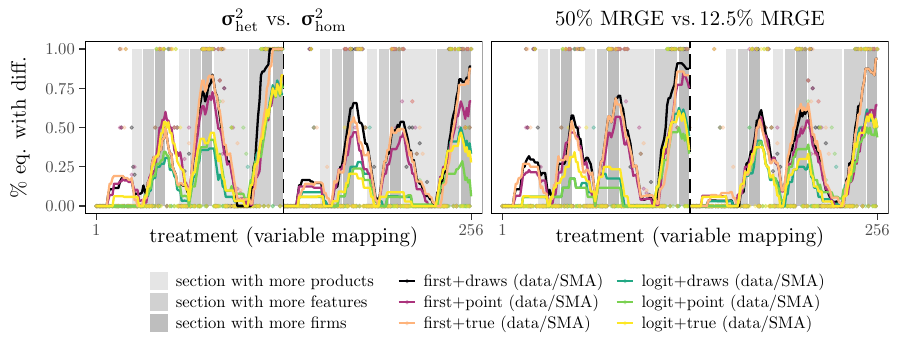} 

}

\caption{Percentage of equilibria having a differentiated product (line) for at least one competitor}\label{fig:difflines}
\end{figure}

\noindent  \textbf{\emph{Differentiation}} \newline \noindent 
\autoref{fig:difflines} presents the percentage of equilibria in which
not all competitors share the same product (line). The flips are
excluded again because they bias the \#equilibria through the number of
differentiated ones as described in the section on preliminary measures.
Similar to the \#equilibria, the shares of differentiated equilibria
based on the estimated parameter sets closely follow the true percentage
within both choice rules, and the disparity between the choice rules,
which manifests as fewer differentiated logit choice equilibria, might
be ascribed to the variance of mutual best responses (lower for logit).
The spikes do not coincide with those in the \#equilibria
(cf.~\autoref{fig:prelimeqmeasures}).\footnote{Even if they did, the
  shares would still remain unchanged unless there was a disproportional
  increase of differentiated equilibria.} Here, the peaks are observable
in competitive scenarios with more firms for both choice rules, which is
plausible, as only one firm has to deviate. In these sections, the SMA
tend to rise further with heterogeneity. The other two imperfect
experimental factors (\#features and \#products) as well as the error
magnitude do not seem to be as influential. The reason for classifying
this measure as pivotal is rooted in the fact that although an identical
share of differentiated equilibria for a true and an estimated parameter
set does not automatically translate into structural equality, it
definitely suggest the latter much more strongly than a congruence in
the preliminary measures.

\begin{figure}[!ht]

{\centering \includegraphics{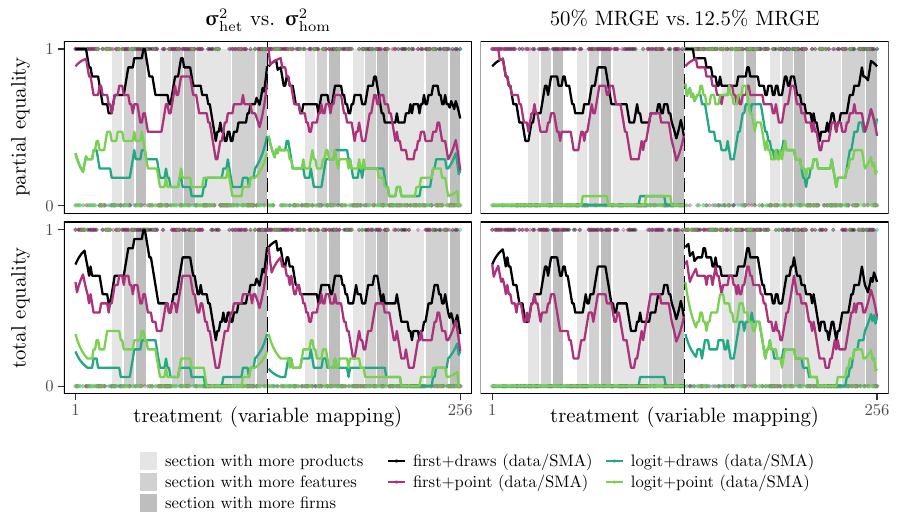} 

}

\caption{Ability to uncover the true first choice equilibria}\label{fig:equalityfirst}
\end{figure}

\begin{figure}[!ht]

{\centering \includegraphics{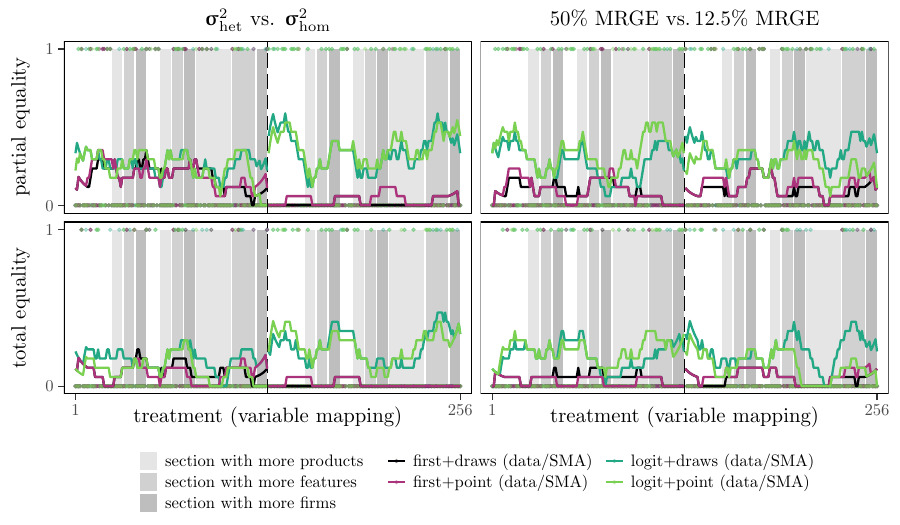} 

}

\caption{Ability to uncover the true logit choice equilibria}\label{fig:equalitylogit}
\end{figure}

\noindent  \textbf{\emph{Ability to uncover the true equilibria}}
\newline \noindent Now, when the ability to reveal the true equilibria
is explicitly examined, the above impression of structural equality
cannot be universally confirmed. In \autoref{fig:equalityfirst} and
\autoref{fig:equalitylogit}, we match the equilibria based on the two
choice rules (first, logit) \(\times\) two estimated parameter sets
(draws, point) against the true first choice and the true logit choice
equilibria, respectively. We speak of total equality if a combination
leads to nothing but the true equilibria and of partial equality if the
true equilibria or the true equilibria plus additional ones are found
(total equality therefore is a subset of partial equality). We
deliberately chose not to compute and display the percentage of
uncovered true equilibria because in reality the management of the
client firm commissioning such an equilibrium simulation is the only one
who can classify the solutions with regard to their usefulness. If we
did, a high detection rate could be deceptive insofar as it would create
an extremely misleading sense of certainty in case the few missing
equilibria were of paramount importance. The flips are not excluded
because the firms' symmetry and the tie-breaking strategy for the first
choice rule (see \hyperref[sec36]{Section 3.6}, not necessary for the
logit choice rule) guarantee that whenever a combination leads to a
differentiated true equilibrium, the corresponding flips always come up
and match too.

\autoref{fig:equalityfirst} refers to the true first choice equilibria
and shows that first+draws generally performs best in uncovering them,
closely followed by first+point. Logit+draws and logit+point do equally
worse here in the majority of cases. This holds true for most
experimental conditions across the other two perfect experimental
factors (preference structure and error magnitude) and their levels as
well as both concepts of equality. The results of the former two
combinations tend to improve further with heterogeneity and the latter
two fail completely when there is a lot of disturbance. Surprisingly,
the first choice SMA \setlist{nolistsep}

\begin{itemize}[noitemsep]
    \item do not fall below the logit choice SMA in sections with more products 
          although having many more equilibria (cf.
          \autoref{fig:prelimeqmeasures}) and
    \item are not overtaken by the logit choice SMA in sections with more firms 
          although having much higher shares of differentiated equilibria (cf. 
          \autoref{fig:difflines}).
  \end{itemize}

\noindent

If both choice rules were similarly stable in the prediction of their
own true equilibria, one would expect the above to be exactly reversed
for the true logit choice equilibria. At first glance,
\autoref{fig:equalitylogit} does not fully reject this hypothesis, but
there are major differences to be elucidated. Firstly, the draws do not
surpass the posterior means as distinctly in either rule. Secondly, the
first choice rule mostly performs worse than the logit choice rule in
\autoref{fig:equalityfirst}. Thirdly, while the first choice SMA still
slightly decreases with homogeneity, the logit choice rule benefits from
it, and lastly, the latter does not really improve in its overall
detection rate compared to \autoref{fig:equalityfirst}, except when the
error is large. It is precisely this steadiness of the logit choice rule
that advocates an instability, even though the performance ranking of
the rules is indeed reversed.

\begin{figure}[!ht]

{\centering \includegraphics{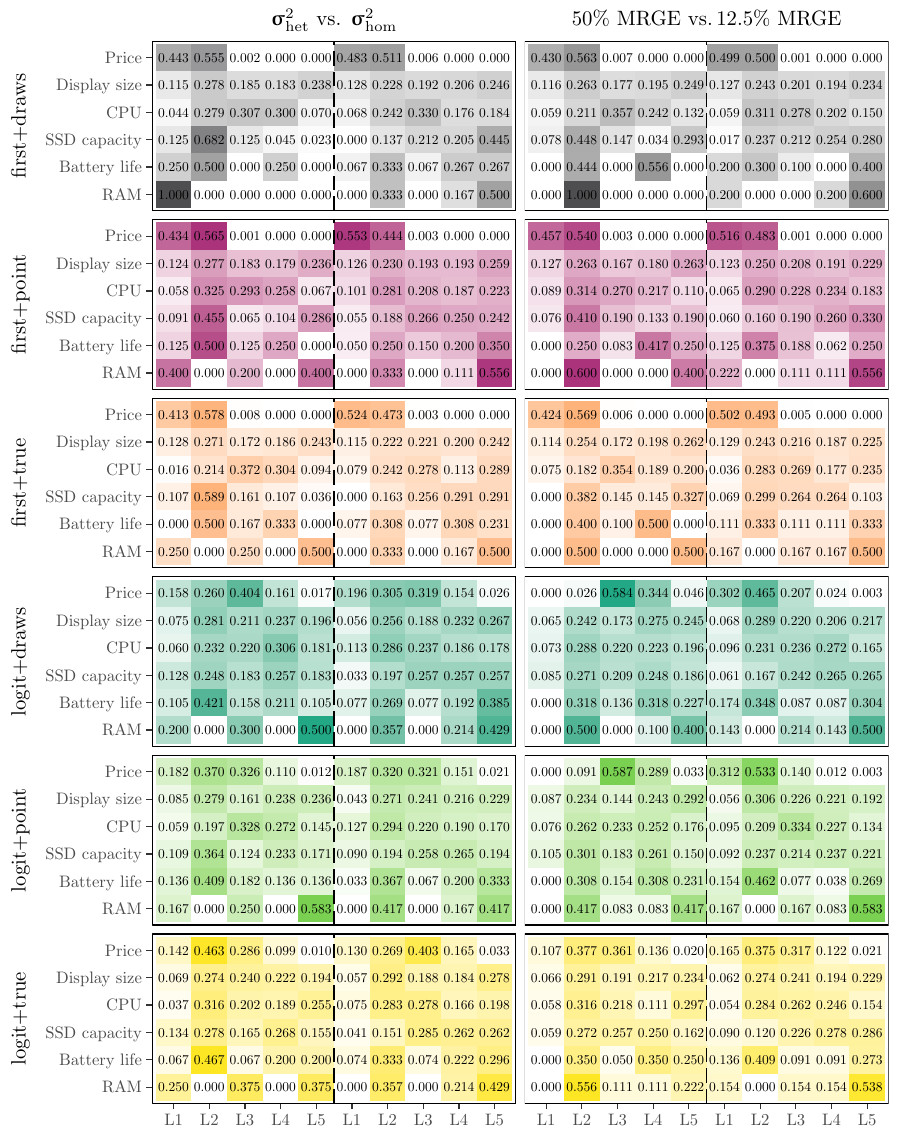} 

}

\caption{Relative level frequencies across all equilibria}\label{fig:lvlfreq}
\end{figure}

To track down the reasons behind this constant but rather weak detection
rate of the logit choice rule, \autoref{fig:lvlfreq} displays the
relative frequency of each level of each feature across all equilibrium
solutions. To facilitate a more detailed analysis of the structure of
the equilibria, we are, for once, forced to somewhat modify our visual
scheme. The six choice rule \(\times\) parameter set combinations are
now separate rows (recognizable through the heatmap colors and the
y-axis titles), and the axes show the features (y) and the levels (x).
Since the imperfect experimental factors can no longer be integrated
implicitly, their separation is intentionally omitted to not compromise
interpretability. We have to be aware that the (relative) level
frequencies are impacted by the variance of \setlist{nolistsep}

\begin{itemize}[noitemsep]
    \item the number of equilibria in a run (recall the spikes of the first 
          choice rule),
    \item the number of products and firms in an equilibrium,
    \item the number and the settings of runs with the respective feature (e.g.,
          RAM is only present in one of 16 base conditions) and
    \item the differentiation of the firms (flips are included as in 
          \autoref{fig:equalityfirst} and \autoref{fig:equalitylogit}).
  \end{itemize}

\noindent If we look at price, we notice that the first choice rule
usually leads to equilibria with lower prices and demonstrates a much
greater stability in price across all parameter sets (first three
differently colored rows in \autoref{fig:lvlfreq}) and the other two
perfect experimental factors (preference structure and error magnitude)
as well as their levels (columns of \autoref{fig:lvlfreq}). This is
crucial, as price is the main driver of contribution margins (i.e.,
equilibria), and already reveals why the logit choice rule stays fairly
untouched by a change of reference (matching against true first or logit
choice equilibria) while the first choice rule transitions from high to
low equilibrium recovery (cf.~\autoref{fig:equalityfirst} and
\autoref{fig:equalitylogit}). Moreover, the most apparent discrepancy in
price, which happens to be between logit+draws/logit+point and
first+true for substantial disturbances (mainly L3 (899 €) and L4 (1,199
€) vs.~L1 (299 €) and L2 (599 €), respectively) explains the concomitant
collapse of the logit choice rule in uncovering the true first choice
equilibria in \autoref{fig:equalityfirst}.

As the assessment of the design (i.e., non-price features) stability is
less straightforward and price is more critical, let us go one step
further and compare the mean absolute error (MAE) of the price and
design feature level frequencies between the true and estimated
parameter sets within each choice rule (\autoref{tab:MAElvlfreqmat}).
The price MAE range of the first choice and logit choice rule is
0.2\%-1.6\% and 3.7\%-18.3\%, respectively, which confirms the
(in)stability. Unsurprisingly, the design MAE is rather high for both
choice rules and lies in between the price MAE ranges (2.9\%-9.8\% for
first, 2.0\%-6.4\% for logit), which is probably due to its inferior
decisive power (i.e., less impact on the contribution margin because of
relatively low costs compared to price) and different amounts of data
per feature. \newline

\begin{figure}[!ht]

{\centering \includegraphics{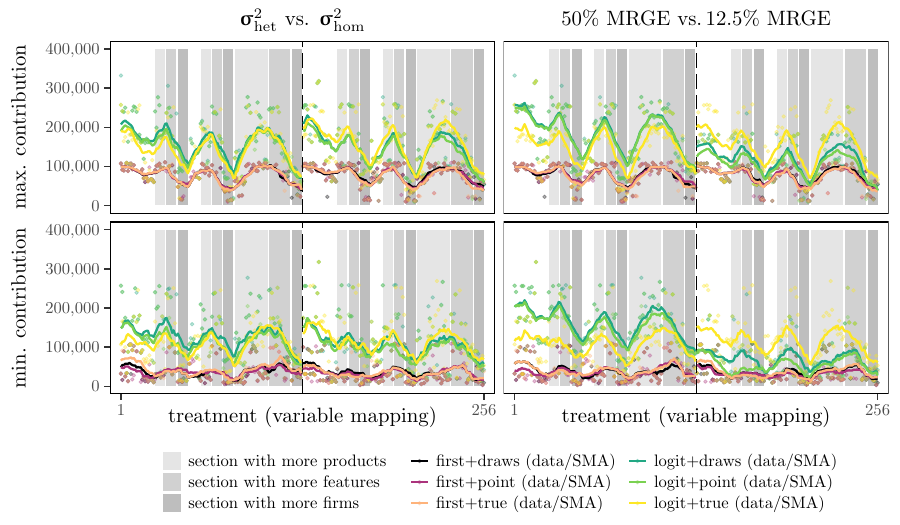} 

}

\caption{Contribution margin bounds}\label{fig:contribution}
\end{figure}

\noindent \textbf{\emph{Contribution margins}} \newline \noindent In the
event of price playing a superior role and equilibria in particular
exhibiting divergent prices depending on the choice rule and the
parameter set, the contribution margins should reflect that. Coming back
to the initial visual scheme, \autoref{fig:contribution} therefore shows
the maximum and the minimum contribution margin across the equilibria of
each experimental condition and confirms the findings from
\autoref{fig:lvlfreq}. Flips are included, as the contribution margins
can be different. While the logit choice SMA for the true and estimated
parameter sets rarely align (result of price instability), the first
choice SMA almost always do (due to price stability) and consistently
stay far below the logit choice SMA (because of lower prices). Do not be
deceived by the proximity of the true and the estimated logit choice SMA
for the levels of preference heterogeneity, as it is likely to be caused
by the mirroring of the estimated values along the true ones for the two
error magnitudes identical to the prices in \autoref{fig:lvlfreq}
(yellow line below and above green lines for 50\% and 12.5\% MRGE,
respectively). The local dips in sections with more firms, which are
comparable for both choice rules on a log scale, are a consequence of
the zero-sum game with respect to share, i.e., the same market volume
has to be apportioned among more firms.

\begin{table}[H]
\centering
\caption{\label{tab:MAElvlfreqmat}Mean absolute error between relative level frequencies}
\centering
\fontsize{9}{11}\selectfont
\begin{tabular}[t]{lcccc}
\toprule
Relative level freq. MAE for & $\bm{\upsigma}^2_\mathrm{het}$ & $\bm{\upsigma}^2_\mathrm{hom}$ & $50\%$ & $12.5\%$\\
\midrule
\addlinespace[0.3em]
\multicolumn{5}{l}{\textit{Price}}\\
\hspace{1em}first+draws vs. first+true & 0.012 & 0.016 & 0.002 & 0.003\\
\hspace{1em}first+point vs. first+true & 0.008 & 0.012 & 0.013 & 0.005\\
\hspace{1em}logit+draws vs. logit+true & 0.081 & 0.041 & 0.183 & 0.091\\
\hspace{1em}logit+point vs. logit+true & 0.037 & 0.044 & 0.157 & 0.122\\
\addlinespace[0.3em]
\multicolumn{5}{l}{\textit{Design}}\\
\hspace{1em}first+draws vs. first+true & 0.098 & 0.029 & 0.068 & 0.048\\
\hspace{1em}first+point vs. first+true & 0.056 & 0.041 & 0.056 & 0.046\\
\hspace{1em}logit+draws vs. logit+true & 0.040 & 0.020 & 0.041 & 0.024\\
\hspace{1em}logit+point vs. logit+true & 0.064 & 0.032 & 0.053 & 0.034\\
\bottomrule
\end{tabular}
\end{table}

\subsection{6 Conclusions}\label{sec6}

While the fields of conjoint and discrete choice analysis as well as
product line optimization have advanced substantially over the past few
decades, and although there is a growing interest in the subsequent
simulation of competitive reactions to support managerial decisions,
comparatively few works concentrated on the computation of
conjoint-based equilibrium solutions. In this paper, we undertook a
large-scale Monte Carlo study in order to address three unresolved
fundamental research questions in this regard. We were curious if
state-of-the-art mixed logit models are even capable of uncovering the
true equilibria arising under the true consumer preferences, how the
structural properties and detection rates vary for different types of
choice behavior, and whether using fully Bayesian choice models (i.e.,
posterior draws instead of means) for simulation is preferable.

Our analysis of thousands of equilibria, derived in full and numerically
exact from the competitive dynamics among multinational computer
manufacturers given real prices and costs, provides evidence that
researchers and practitioners who are concerned with simulating Nash
equilibria for product design based on conjoint choice frameworks
primarily have to choose which choice rule more realistically models the
decision making of their target group. Irrespective of the number of
features, products per line and firms in the market, as well as the
degree of preference heterogeneity and disturbance, competitive
reactions should be simulated by applying the first choice rule to
Bayesian posterior draws (if computationally feasible) in case of more
deterministic consumer behavior and the logit choice rule to posterior
means otherwise to optimize the recovery of the true equilibria. In the
former setting, however, the detection rate is likely to be considerably
higher (also for posterior means). It is imperative that the choice rule
premise is only motivated by the expected behavior and not the remaining
findings of this paper (increased product differentiation and price
stability for the first choice rule, increased prices and contribution
margins for the logit choice rule), as they are consequences of the
assumed truth.

The limitations of our study are congruent with potential avenues for
future research. It would be worthwhile to ascertain whether the results
obtained in this paper can be replicated in case of (1) asymmetric
competitors (in, e.g., price, cost structure, number of products), (2) a
superior decisive power of the cost structure (i.e., greater impact of
the design (non-price) features on the contribution margin), (3) more
advanced optimization constraints like reduced manufacturing costs
through shared feature levels in a product line
(\citeproc{ref-2009_Wang_et_al}{Wang et al., 2009}), (4) segment
heterogeneity (and methods explicitly capturing such preference
structures, as finite mixture models, see, e.g.,
\citeproc{ref-2002a_Andrews_et_al}{Andrews et al., 2002a},
\citeproc{ref-2002b_Andrews_et_al}{2002b};
\citeproc{ref-2024_Goeken_et_al}{Goeken et al., 2024}), (5) the
Stackelberg equilibrium concept (see, e.g.,
\citeproc{ref-2010_Steiner}{Steiner, 2010}), (6) the integration of a
no-choice option (see \hyperref[sec37]{Section 3.7}), and (7) omitted
monotonicity constraints.

Finally, the development of an approach for pruning the vast initial
state space (see \hyperref[sec37]{Section 3.7}) without loss of
equilibrium information would be a significant milestone, as the
simulation of more complex scenarios might thereby come within reach.

\subsection{Acknowledgments}\label{acknowledgments}

The authors thank Philipp Aschersleben for his helpful comments on the
pre-optimization described in \hyperref[sec37]{Section 3.7}.

\subsection{References}\label{sec7}

\bibliography{}

\phantomsection\label{refs}

\end{document}